# Source apportionment of air pollution burden using geometric non-negative matrix factorization and high-throughput multi-pollutant air sensor data in Curtis Bay, Baltimore, USA

*Bora Jin*+, Bonita D. Salmerón‡, David McClosky¢, David H. Hagan¢, Russell R. Dickerson¥, Nicholas J. Spadaº, Lauren N. Deanes‡, Matthew A. Aubourg‡, Tarunika Ramprasad₹, Laura E. Schmidt‡, Gregory G. Sawtell§, Christopher D. Heaney*‡¤#, Abhirup Datta*†*

+Department of Applied Mathematics and Statistics, Colorado School of Mines, Golden, CO, 80401, USA

‡Community Science and Innovation for Environmental Justice Initiative, Center for a Livable Future, Department of Environmental Health and Engineering, Johns Hopkins Bloomberg School of Public Health, Baltimore, MD, 21205, USA

¤Department of Epidemiology, Johns Hopkins Bloomberg School of Public Health, Baltimore, MD, 21205, USA

#Department of International Health, Johns Hopkins Bloomberg School of Public Health, Baltimore, MD, 21205, USA

¢QuantAQ, Inc, Somerville, MA, 02143, USA

¥Department of Atmospheric and Oceanic Science, University of Maryland, College Park, MD, 20742, USA

ºAir Quality Research Center, University of California, Davis, CA, 95616, USA

₹Materials Characterization and Processing, Johns Hopkins Whiting School of Engineering, Baltimore, MD, 21211, USA

§Community of Curtis Bay Association, Baltimore, MD, 21226, USA and South Baltimore Community Land Trust, Baltimore, MD, 21225, USA

†Department of Biostatistics, Johns Hopkins Bloomberg School of Public Health, Baltimore, MD, 21205, USA

ABSTRACT

Air sensor networks provide hyperlocal, high-frequency data on multiple pollutants, but unlike speciated particulate matter (PM) measurements, they lack direct chemical signatures for source identification. However, these networks collect data at high temporal resolution and multiple spatial locations, creating new opportunities to interpret latent factors through their relationships with spatial proximity to known origins, temporal patterns, and meteorology. We analyze 451,946 one-minute air sensor records from Curtis Bay (Baltimore, USA; October 2022 – June 2023), covering size-resolved PM, black carbon (BC), carbon monoxide (CO), nitric oxide (NO), and nitrogen dioxide ($NO_2$), using a geometric non-negative matrix factorization (NMF) approach that scales to large datasets and yields provably unique factor attribution percentages. Three stable latent factors emerge with converging evidence toward recognizable source categories: Factor 1 explains > 70% of fine and coarse PM and ~30% of BC; Factor 2 dominates CO and contributes ~70% of BC, NO, and $NO_2$; Factor 3 is specific to the larger PM fractions, $PM_{10}$ to $PM_{40}$. Regression analyses of the estimated factor intensities with wind direction and bulldozer activity at a nearby coal terminal and a case study on a known bulldozer incident link Factors 1 and 3 to the coal terminal. Extreme-intensity episodes from Factors 1 and 3 averaged ~33 and ~24 minutes per day at the site nearest the coal terminal, attenuating with distance. Factor 2 reflects diurnal traffic patterns. Together, these results show that dense air sensor networks paired with the geometric NMF method can move community air monitoring beyond pollution detection toward identifying likely source categories and informing actionable mitigation strategies.

SYNOPSIS

For decades, Curtis Bay community residents have reported the accumulation of dark dust at and in their homes, which many attribute to the operation of a neighboring open-air coal terminal. This study uses geometric non-negative matrix factorization to transform hyperlocal air sensor data into scientific evidence about this longstanding concern, while also demonstrating its utility as a reliable source apportionment framework for community air monitoring and regulatory decision making.

## Introduction

Air pollutants in communities arise from multiple origins, and effective policy depends on knowing which sources or source categories dominate each pollutant. Source apportionment analysis addresses this need by quantifying contributions of latent factors to observed concentrations. Multivariate receptor modeling, a popular class of source apportionment approaches, is often posed as a non-negative matrix factorization (NMF) problem, in which the data matrix is decomposed into two factor matrices: a factor intensity matrix and a per-unit factor profile matrix. Some source apportionment methods (e.g., chemical mass balance[1,2]) assume the factor profile matrix is known, in which case estimating factor intensities reduces to a regression problem. Other methods, such as positive matrix factorization (PMF)[3] and generic NMF software (in Python or R), treat both matrices as unknown and are thus closer to factor models. These methods typically estimate the unknown factor matrices by solving a (weighted) least-squares optimization problem[4,5]. More specialized extensions accommodate temporal[6] and spatial[7] correlation across multi-pollutant data and admit Bayesian formulations[8,9].

NMF is generally non-unique, and the factor matrices are not individually identifiable, unless imposing additional assumptions. The Environmental Protection Agency's (EPA) documentation of PMF[10] explicitly acknowledges this non-identifiability and provides rotation controls such as Fpeak to explore either sparser or more diffuse solutions. PMF also allows external knowledge to be incorporated via constraints, which steer a selected profile or intensity toward scientifically meaningful reference values. These tools can help stabilize latent factors and improve interpretability. Alongside PMF, NMF variants impose explicit sparsity in either of the factor matrices to reduce factor overlap and improve identifiability and estimate the matrices via alternating least squares with regularization penalties[11]. However, we found that their results are often highly sensitive to regularization and other modeling choices.

The recent rise in usage of multi-pollutant air sensor networks[12,13] for exposure assessment presents new opportunity to conduct source apportionment studies at locally-relevant spatiotemporal scales (i.e., micro to meso-gamma meteorological scales). These sensors offer spatially and temporally dense data which can aid in understanding major pollution contributors of communities. However, the high temporal resolution (second- or minute-level) data from air sensors leads to large datasets that pose a scalability challenge to some existing source apportionment tools. For instance, the current EPA PMF implementation is limited to data sets with fewer than 500,000 observations (samples × pollutants)[14], and the practical limit is often lower.

As an alternative to algebraic (least-squares-based) NMF, geometric approaches to NMF have a long and well-established history across scientific applications. These approaches view the data matrix as a point cloud within a conical hull, where the corners of the cloud become factor profiles after suitable scaling. Conditions for identifiability of geometric NMF are formalized[15], and many algorithms are available for leveraging convex hull geometry. It has found particularly successful use in hyperspectral unmixing, where it is known as vertex component analysis or endmember extraction[16–18]. Despite this maturity, its adoption in air pollution source apportionment has been limited. While an early work, Unmix[19], drew the connection, existing implementations are not compatible with modern computing environments, leaving an established approach largely untapped in this domain.

Recent work on geometric NMF (GeomNMF)[20] has demonstrated how geometric ideas can lead to provably reliable and scalable source apportionment. GeomNMF focuses on an identifiable quantity, the factor attribution matrix, which is the percentage of concentrations of each pollutant attributable to each latent factor. This matrix is shown to be unique, scale-invariant, robust to separable measurement errors, and interpretable, even when individual factor matrices in NMF are not. Under weaker and more realistic assumptions than those typically required for factor identifiability, GeomNMF provides a scalable geometric approach to estimating this factor attribution matrix along with accuracy guarantees. Building on this foundation, the aim of this manuscript is to apply GeomNMF for source apportionment in Curtis Bay (Baltimore, USA) using a multi-pollutant, hyperlocal air sensing network dataset with nearly half-a-million samples, and to use the inferred factors to address community environmental concerns.

Curtis Bay is a neighborhood where a dense residential community is situated near a heavy concentration of traffic and industrial emissions[21]. One facility in particular, an open-air coal export terminal operated by CSX Corporation, has long been associated with community environmental health concerns related to air pollution burden[21,22] and fugitive dark dust (shown to contain coal dust[23]) at homes, businesses, schools, and churches in the area. Community members report visible pollution events that are episodic and spatially concentrated near the terminal[24], with elevated coarse-mode aerosol concentrations consistent with local fugitive dust rather than long-range regional transport or secondary chemical formation.

To characterize air pollution in Curtis Bay, a network of sensors has been deployed across multiple monitoring sites, measuring several pollutants at one-minute resolution. Source apportionment from such networks has distinct characteristics relative to traditional source apportionment. Traditional source apportionment in the Baltimore area, for instance, has relied on speciated particulate matter (PM) measurements at a single monitoring site[25,26], which limits spatial resolution but enables detailed chemical fingerprinting of pollution origins. In contrast, air sensor deployments[27,28] typically cannot speciate aerosol composition, but provide information on particle size, particle number, and gas concentrations at high spatiotemporal resolution. Air sensor data are increasingly common in community air quality efforts and have supported a range of source apportionment applications: clustering approaches have demonstrated their capability for source differentiation[29]; NMF has effectively separated combustion from non-combustion particle sources[12,30]; and k-means clustering and PMF have been used to identify sources in real-world scenarios and quantify effects of activities from construction sites, roadsides, and quarries[31]. Building on this growing literature, we apply geometric NMF to extract interpretable evidence about likely contributing source categories from high-resolution monitoring data of the Curtis Bay sensor network. Throughout, estimated factors reflect pollutant mixtures tied to source categories rather than individual emission origins.

High-frequency air sensing data is particularly well suited to geometric NMF. Geometrically, the large sample size allows the observed pollutant point cloud to more fully populate the enclosing cone, improving estimation of factor profiles as corners of this cone. This in turn sharpens estimation of the factor intensities and any subsequent quantities such as the factor attribution matrix. The spatiotemporal resolution of the data further enables factor intensity estimates to be interpreted through their spatial and temporal patterns, with formal downstream analyses regressing these intensities on other spatial-temporal variables (e.g., being downwind from the terminal or ongoing activity in the terminal) providing triangulating evidence.

Within this framing, source apportionment is well positioned to assess the Curtis Bay pollution landscape. The geometric NMF analysis scales to and benefits from the high-volume dataset, providing precise estimates of factor attributions and (up to scaling) profiles and intensities, from pollutant data alone, without relying on prior information about nearby facilities. Given the global prevalence of mixed urban/industrial areas and the increasing adoption of air sensor networks, this study offers a widely applicable pipeline for factor attribution in community settings to inform policy, regulation, and mitigation.

## Materials and Methods

### Data

Data were collected in Curtis Bay through a community-driven collaboration among the Community of Curtis Bay Association (CCBA), the South Baltimore Community Land Trust (SBCLT), academic partners, and the Maryland Department of the Environment (MDE) to evaluate the potential impacts of the coal terminal on Curtis Bay air pollution burden; details of the collaboration are available elsewhere[23].

The coal terminal in Curtis Bay can handle up to 14 million tons of coal per year, making Baltimore City one of the major coal export hubs in the United States[32]. This facility is designed for large-scale storage, loading, and unloading. Coal handling involves a sequence of activities, including dumping coal from trains, transporting it via conveyors to transfer towers, and loading it onto ships or barges, with bulldozers widening coal piles or pushing coal to conveyor inlets[33](see the left panel of Figure 1).

We deployed QuantAQ MODULAIR monitors (QuantAQ Inc., Somerville, MA) across 10 monitoring sites in Curtis Bay to measure PM concentrations: $PM_1$, $PM_{2.5}$, $PM_{10}$, total suspended particles (TSP; equivalent to $PM_{40}$) in μg/m$^3$; gases: carbon monoxide (CO), nitric oxide (NO), and nitrogen dioxide ($NO_2$) in ppb; and temperature in degrees Celsius. At four of these sites (Locations 1, 2, 5, and 8; map provided on the right panel of Figure 1), we also deployed Distributed Sensing Technologies (DSTech) ObservAir sensors to measure black carbon (BC) in $\mu g/m^3$. Analysis is restricted to these four sites, where the full pollutant suite including BC was measured. Prior work has reported high agreement between regulatory-grade measurements and MODULAIR units[22,34,35] and DSTech ObservAir monitors[21] (see Supporting Information S1 for details). Based on this validation, no further adjustments were applied to the PM and gas measurements.

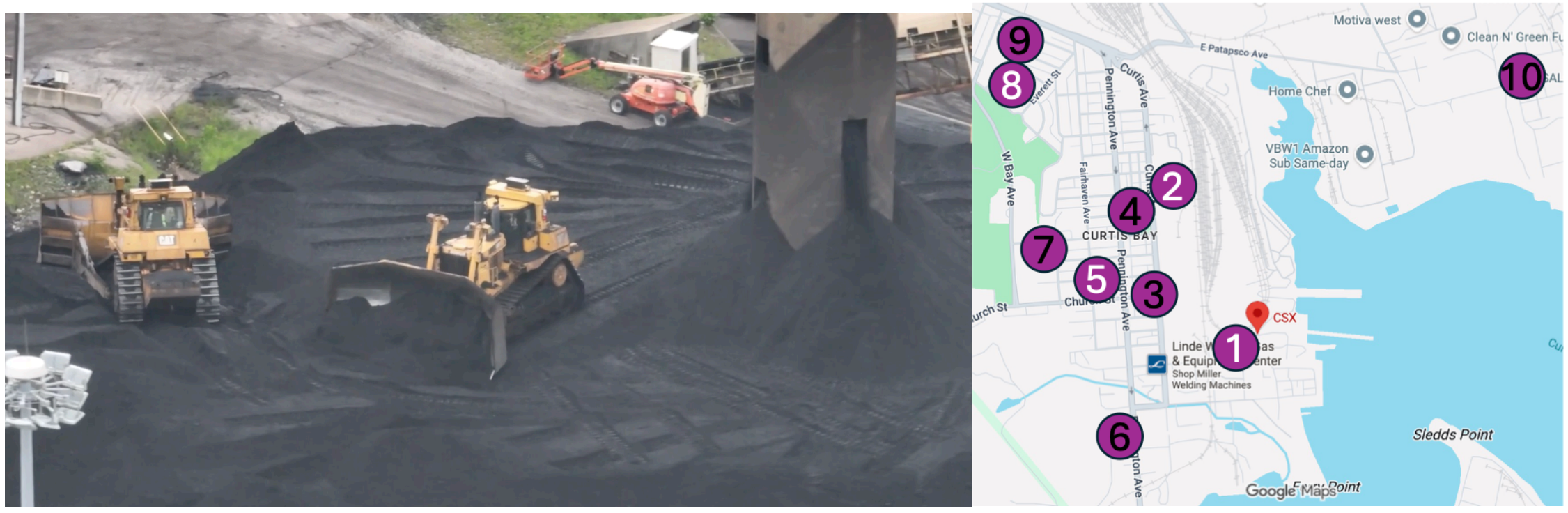

**Figure 1.** Left: Bulldozers working one of the coal piles at the Curtis Bay open-air coal terminal. Right: Google Maps view of the Curtis Bay study area showing ten monitoring sites (purple circles, numbered 1-10) and the coal terminal (red pin). The four monitoring sites analyzed in this manuscript (Locations 1, 2, 5, and 8) are highlighted in white.

Each MODULAIR was paired with a sonic anemometer (Davis Instruments Corp., Hayward, CA) to record wind speed in miles per hour and direction in degrees. A binary downwind indicator was constructed for each site relative to the coal terminal based on measured wind direction. All MODULAIR and ObservAir measurements were collected at 1-minute resolution. Trail-camera footage (CamPark Electronics Co., Ltd, Hong Kong) near Location 5 was used to construct a binary minute-level time-series of visible bulldozer activity. Because bulldozer operation is the most readily observable component of coal handling, we use it as a partial proxy for the broader coal handling process at the terminal. Hourly solar radiation in $W/m^2$ from the MDE's Essex station (~15 km from Curtis Bay) was mapped to the corresponding minute-level samples. We applied previously described data processing and quality assurance/quality control procedures and refer readers to the same source for information on the monitoring locations and the detailed definition of the downwind variable[22]. To study association of estimated factor intensities with traffic, minute-level data on traffic count and classification were collected using the Houston Radar Armadillo Tracker (Houston Radar LLC, Sugarland, TX) on two parallel routes in Curtis Bay. Further details are provided in Supporting Information S1.

For source apportionment, we first formed non-overlapping bins in terms of PM size increments: $PM_1$, $PM_{2.5}$-$PM_1$, $PM_{10}$-$PM_{2.5}$, TSP-$PM_{10}$ to avoid double counting across size bins. Sensitivity to binning choices is conducted as detailed later. We retained complete-case minute-level records with all the following measurements available: $PM_1$, $PM_{2.5}$-$PM_1$, $PM_{10}$-$PM_{2.5}$, TSP-$PM_{10}$, BC, CO, NO, and $NO_2$. We additionally removed two clear outliers in TSP-$PM_{10}$, yielding a total of 451,946 minute-level records (or 3,615,568 total observations) spanning 11:21 am on October 21, 2022, to 11:59 pm on June 16, 2023. Meteorological and coal terminal activity variables may have fewer observations, but they are not used in the main source apportionment analysis, only for subsequent analyses on interpreting the estimated factors.

Infrequently (45,039 records; < 10% of data), BC values are negative, which has been described in detail previously[22]. Because source apportionment relies on NMF, and thus requires non-negative inputs, we apply an offset transform that preserves ranks while making all values non-negative. We then apply quantile-based scaling which respects each pollutant's measured range, avoiding the over-compression that occurs when pollutants with heavy tails are normalized to a fixed range (e.g., [0,1]). For pollutant $j$, let $m_j$ be the minimum and $q_{0.85,j}$ be the 85th percentile.

We transform $Y_{ij}^0$, the measured raw concentration of pollutant $j$ in sample $i$, to $Y_{ij} = \{Y_{ij}^0 - \min(m_j, 0)\}/(q_{0.85,j} - m_j)$. The mean shift is applied only when $m_j < 0$, which occurs only for BC, as described above. The scaling by $q_{0.85,j} - m_j$ provides outlier-robust normalization. We use the 85th percentile because all pollutants are highly skewed. For example, TSP-$PM_{10}$ is so skewed that its values are mostly zero below the 85th percentile. Sensitivity to the choice of the 85th percentile is conducted as detailed later.

**Geometric source apportionment**

In source apportionment, the observed (possibly normalized) data matrix of concentrations $Y$ is modeled as the sum of contributions from a few latent factors. With $K$ factors, this yields an NMF representation $Y = WH$, where $Y$ is an $n \times J$ non-negative matrix of pollutant measurements for $n$ samples across $J$ pollutants, $W$ is an $n \times K$ non-negative matrix of $K$ factor intensities across samples, and $H$ is a $K \times J$ non-negative matrix whose rows represent the per-unit pollutant profile of each factor. Since only $Y$ is observed, the factor matrices $W$ and $H$ are not individually identifiable due to scale ambiguity.

Geometric NMF approaches view each record of the data $Y$ as a point in a conical hull, or equivalently, each row of the row-normalized data $Y^*$ as a point in a convex hull. More data points allow a better understanding of the shape of the point cloud, leading to a more accurate estimate of the corners of the convex hull, which are simply the rows of $H^*$, the row-normalized version of $H$. This yields an equivalent NMF representation $Y = \widetilde{W} H^*$, where $\widetilde{W}$ absorbs the row scales of $H$ and is paired with $H^*$.

The scale ambiguity means that $(W, H)$ and $(\widetilde{W}, H^*)$ are equally valid representations, making inference on individual factor matrices ill-posed. The GeomNMF framework[20] addresses this by targeting a quantity that is unaffected by this scale indeterminacy and thus identifiable across different NMF representations: the factor attribution matrix $\Phi = (\phi_{kj})$, defined as

$$\phi_{kj} = \frac{\tilde{\mu}_k H^*_{kj}}{\sum_{l=1}^{K} \tilde{\mu}_l H^*_{lj}}, \qquad \tilde{\mu}_k = \mathbb{E}(\widetilde{W}_{ik}), \tag{1}$$

where $\tilde{\mu}_k$is the expected (statistical average) intensity for factor $k$. Each element $\phi_{kj}$ is the fraction of pollutant $j$ attributable to factor $k$ and by construction $\sum_{k=1}^{K} \phi_{kj} = 1$ for each pollutant $j$, consistent with the intuition that the fractions of a pollutant attributable to all factors must sum to one. Because it is defined in terms of expected rather than observed intensities, $\Phi$ is stable across datasets. Importantly, $\Phi$ also stays the same under multiplicative separable measurement errors of the form $Y'_{ij} = \epsilon_{ij} Y_{ij}$ with $\epsilon_{ij} = a_i b_j$, where $a_i$ captures a sample-specific scaling such as meteorology-driven dilution and $b_j$ captures a pollutant-specific scaling such as instrumental sensitivity.

Given a specified number of factors $K$, GeomNMF estimates $H^*$ by first computing the sample convex hull of $Y^*$, then selecting the $K$-point subset of hull corners that maximizes the volume of the resulting polytope. These points form the rows of $\widehat{H}^*$. This geometric intuition is illustrated schematically in Figure 2. Subsequently, $\widetilde{W}$ is estimated using a constrained linear solve $Y \approx \widetilde{W}\widehat{H}^*$, and $\Phi$ is computed from the estimated factor matrices using (1). For algorithmic details, see Algorithm 1 of the original paper[20].

The matrices $H^*, \widetilde{W}$, and $\Phi$ can be provably accurately estimated using large datasets under realistic conditions. Specifically, the conditions allow standard spatiotemporal dependence in factor intensities and require only a probabilistic separability condition: that each factor occasionally dominates the mixture in a probabilistic sense. This is more plausible than deterministic separability or sparsity assumed in other geometric NMF approaches, which unrealistically demands complete absence of other factors. As a consequence of the scale ambiguity inherent to NMF, estimated intensities in $\widetilde{W}$ are meaningful only in relative terms: they can be compared across time or space within a factor, but not in absolute terms or across factors. Our downstream analyses are designed accordingly.

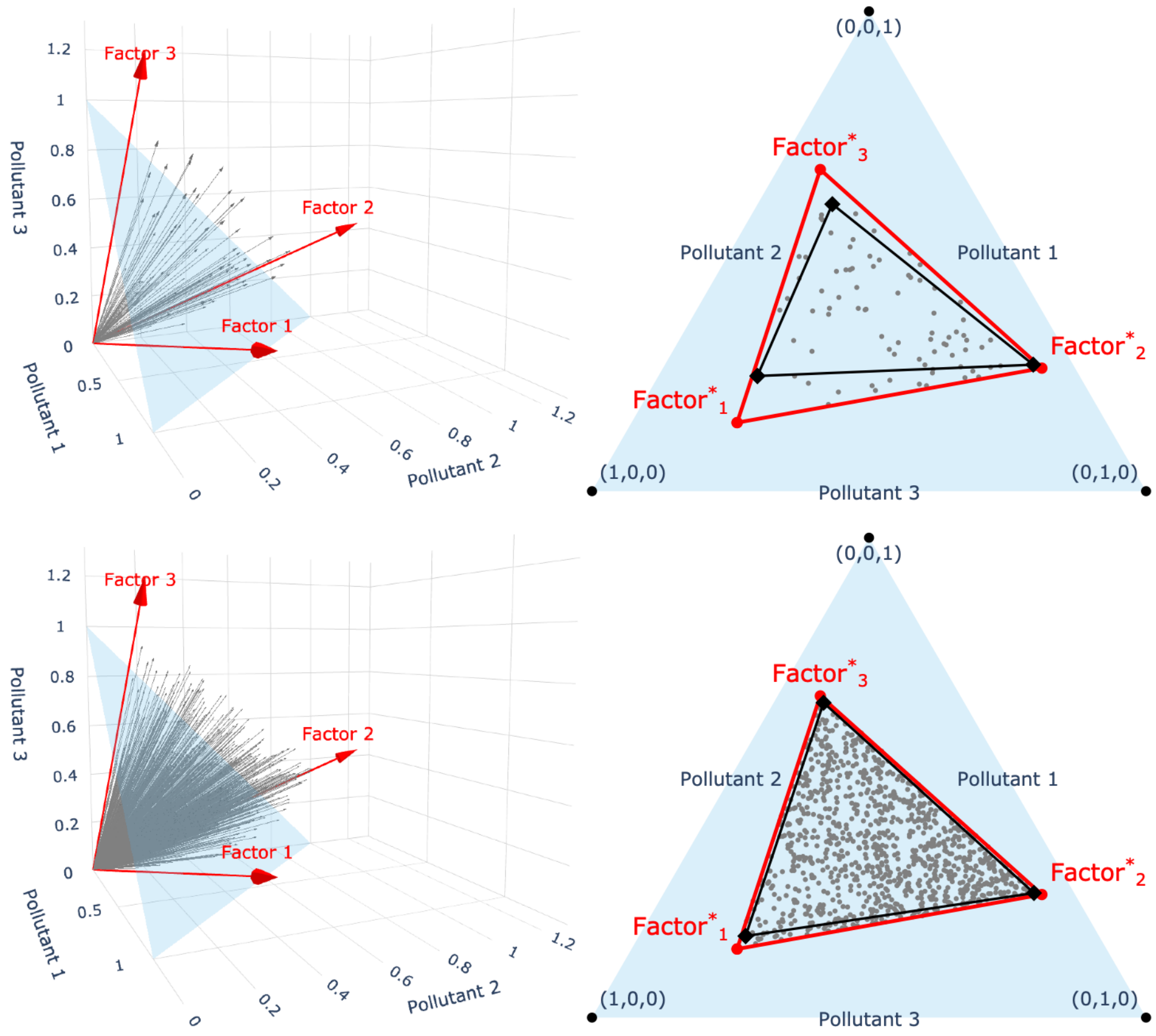

**Figure 2.** Illustration of GeomNMF with three pollutants and three factors. The left panels show original data (gray vectors) together with the rows of $H$ (red vectors; Factors 1–3). The data lie in the unbounded triangular cone (the conical hull) spanned by the rows of $H$. Row normalization is equivalent to projecting the data onto the canonical simplex (blue plane); the corresponding projections are shown in the right panels. Each gray point is a row-normalized observation $Y^*$, and the rows of $H^*$ appear as vertices (Factors* 1–3) of the red triangle (the convex hull). Given the data, the three vertices that maximize their triangle area are shown as black diamonds (i.e., the rows of $\widehat{H}^*$). As the sample size increases (top: n = 80; bottom: n = 1000), the black diamonds approach the red vertices, and thus the estimation accuracy increases.

We quantify uncertainty in $\Phi$ by repeating the estimation over 100 bootstrap resamples of the data. We select the number of factors $K$ based on estimation stability, which also reflects physical interpretability. Specifically, we summarize two metrics of stability: (i) the mean cellwise coefficient of variation of the factor attributions $\phi_{kj}$, computed as the bootstrap standard error (SE) divided by the bootstrap mean and averaged across cells after excluding those with bootstrap mean less than 0.03 to avoid numerical blow-ups; lower is better, and (ii) rank stability of the dominant factor, defined for each pollutant as the bootstrap probability that the top factor (identified from the bootstrap mean) is preserved, averaged across pollutants; higher is better. We report results for $K \in \{3,4,5\}$ factors. We do not consider more factors as the number of stably extractable factors is inherently limited by $J = 8$ measured pollutants.

**Factor interpretation**

The estimated factor attribution matrix $\Phi$ quantifies the contribution of each factor to each pollutant. For non-speciated air sensor data, factor attributions offer a more informative pathway to factor interpretation than factor profiles: attributions can be directly compared against background knowledge of typical source categories for pollutants, whereas profiles require chemical fingerprinting to be meaningful, which is difficult when the variables mix size-resolved particulates and gases and are scaled before factorization. Complementing this, the hyperlocal, high-frequency structure of the air sensor data enables further physical interpretation through converging evidence across spatial, temporal, and meteorological analyses of the estimated columns of factor intensities $\widetilde{W}$. We compute the hourly averages of each estimated factor and compare their diurnal patterns with those of regional traffic counts in Curtis Bay to aid interpretation. We also regress each column of the estimated $\widetilde{W}$ on indicators for bulldozer activity at the coal terminal ("Bulldozer=1"), being downwind of the terminal ("Downwind=1"), and their interaction, adjusting for NO, temperature, wind speed, time of day (AM/PM), solar radiation, and location (Location 1/2/5/8). NO serves as a proxy for local combustion activity[22], and location indicators as time-invariant fixed effects[36]. We focus on the coefficients for "Bulldozer=1", "Downwind=1", and their main effect plus interaction. To validate the factor-level coefficients, we fit the same regression models to each pollutant and assess concordance in the estimated direction of the regression coefficients for pollutants and those of the estimated factors. For gas models, NO is excluded as a covariate. We repeat this for each bootstrap replicate to obtain uncertainty estimates of the regression coefficients.

We further present a case study examining the behavior of the estimated factor intensity during a reported coal terminal incident involving a bulldozer immobilized on a coal pile on February 3, 2023, from 11:30 am to 2:45 pm local time. Recovery efforts to free the bulldozer included repeated track rotation of the bulldozer and temporary deactivation of water sprayers normally used to suppress dust and limit airborne particle emissions, likely resulting in elevated emissions above typical operating levels. To assess the effect of the incident, we compare the distribution of estimated factor intensity during the incident period against a control period consisting of the same hours on the preceding and following days. Distributional differences are summarized via percentiles across 100 bootstrap resamples. We also conduct one-sided two-sample tests: the Mann-Whitney U (MW) test and the Kolmogorov-Smirnov (KS) as descriptive comparisons. For the MW test, the null hypothesis is that the incident and control distributions are identical, and the alternative is that the incident distribution is stochastically larger than the control distribution. For the KS test, the null is that the incident distribution is stochastically smaller than or equal to the control distribution, and the alternative is that this null does not hold. Both tests are nonparametric and accommodate unequal sample sizes.

The spatiotemporal dimension of the air sensing network allows us to address community questions about extreme pollution exposure in Curtis Bay. We summarize factor intensities above factor-specific 90th, 95th, and 99th percentile thresholds, characterizing exceedance frequency, duration, and spatial attenuation across monitoring sites. Formal definitions of all exceedance metrics are provided in Supporting Information S1.

**Sensitivity analysis**

To assess robustness of GeomNMF results, we attempt to fit alternative methods including XRAY[37], least-squares NMF, PMF, and Environmental Source Apportionment Toolkit (ESAT)[38] and compare results where available. XRAY is an alternative geometric NMF method that targets convex hull vertices iteratively, selecting at each step the point that maximally expands the hull boundary. For least-squares NMF, we use the widely used Python package scikit-learn and compare how the estimated attribution matrix $\Phi$ changes under the default (no regularization) setting versus an L2-penalization setting. PMF and ESAT could not be compared due to convergence and scalability limitations as described in Supporting Information S1.

We evaluate sensitivity of the factor attribution matrix $\Phi$ from GeomNMF to measurement uncertainty by propagating the reported uncertainties for MODULAIR[34,39] and ObservAir[40]: 14% for $PM_1$, 11% for $PM_{2.5}$- $PM_1$, 34% for $PM_{10}$- $PM_{2.5}$, 5% for BC, 20% for CO, 20% for NO, and 35% for $NO_2$. Uncertainty for TSP is not yet available, so we use the uncertainty for the next coarsest bin, i.e., $PM_{10}$- $PM_{2.5}$, as an uncertainty estimate for TSP- $PM_{10}$. We generated 100 perturbed datasets by sampling plausible realizations of the latent true concentrations consistent with each pollutant's uncertainty. For example, for $PM_1$ we independently draw $u_{ij} \sim \text{Unif}(-0.14, 0.14)$ and scale the observed $Y_{ij}$ by $1 + u_{ij}$. As a complementary analysis, we assess whether high-uncertainty pollutants destabilize the results. We refit GeomNMF after excluding the two pollutants with the highest reported uncertainty ($NO_2$: 35% and $PM_{10}$- $PM_{2.5}$: 34%) and compare the resulting $\Phi$ for the remaining six pollutants.

We conduct additional sensitivity analyses evaluating robustness to three choices: PM size fraction definitions, by regrouping PM variables as either $PM_1$ and TSP-$PM_1$, or $PM_{2.5}$ and TSP-$PM_{2.5}$; the normalization percentile, by using the $90^{th}$ and $95^{th}$ percentiles in place of the $85^{th}$; and extreme values, by excluding the 30 largest observations for each pollutant (225 samples total, 0.05% of the data). Data and code for analyses presented in this manuscript are publicly available at https://github.com/jinbora0720/GeomNMF-CurtisBay.

## Results

Summary statistics (count, mean, standard deviation, median, minimum, maximum, $15^{th}$, and $85^{th}$ percentiles) of air pollutants (PM, BC, and gases) in Curtis Bay are reported in Table 1. We see that the maximum values of certain pollutants are orders of magnitude higher than even their $85^{th}$ quantiles. For example, the maximum measured concentration of TSP-$PM_{10}$ (17503.5 $\mu g/m^3$) is over 1000 times larger than the $85^{th}$ percentile (14.2 $\mu g/m^3$). Similar phenomenon, albeit to a lesser extent, occurs with most of the other pollutants. Because of such extreme skewness, it is not advisable to scale the data using the maximum values, which leads to inappropriate compressing of the information, and we scale using the $85^{th}$ percentile. A pairwise scatterplot matrix and pollutant-specific normalized histograms appear in Figure S1 in Supporting Information.

**Table 1.** Summary statistics of pollutants measured at 1-minute time interval in Curtis Bay.

| | **Count** | **Mean (Std)** | **Min** | **$15^{th}$ percentile** | **Median** | **$85^{th}$ percentile** | **Max** |
|---|---|---|---|---|---|---|---|
| $PM_1$ (μg/m$^3$) | 451946 | 8.4 (6.3) | 0.09 | 3.2 | 7.0 | 13.7 | 417.0 |
| $PM_{2.5}$ – $PM_1$ (μg/m$^3$) | 451946 | 0.9 (1.4) | 0 | 0.2 | 0.6 | 1.6 | 119.7 |
| $PM_{10}$ – $PM_{2.5}$ (μg/m$^3$) | 451946 | 14.9 (36.6) | 0 | 0.8 | 6.5 | 26.1 | 5233.7 |
| TSP – $PM_{10}$ (μg/m$^3$) | 451946 | 10.7 (148.8) | 0 | 0 | 0 | 14.2 | 17503.5 |
| BC (μg/m$^3$) | 451946 | 0.8 (1.3) | -10.71 | 0.1 | 0.6 | 1.5 | 111.8 |
| CO (ppb) | 451946 | 277.7 (178.7) | 0.05 | 179.9 | 231.1 | 373.3 | 11313.5 |
| NO (ppb) | 451946 | 8.2 (15.4) | <0.001 | 1.7 | 3.2 | 11.9 | 320.4 |
| $NO_2$ (ppb) | 451946 | 13.6 (9.1) | <0.001 | 5.0 | 11.1 | 24.5 | 99.3 |

Figure 3 presents the main source apportionment results with $K = 3$ factors, reporting the mean estimates and bootstrapped SEs of the factor attribution matrix $\Phi$ (numerical values are reported in Table S1, Supporting Information). Factor 1 accounts for most of $PM_{2.5}$-$PM_1$ and around 70% of $PM_1$ and $PM_{10}$-$PM_{2.5}$, with 20% of NO and 30% of $NO_2$. It also contributes around 30% of BC, suggesting that a non-negligible fraction of BC co-occurs with fine to coarse PM. Factor 3 is specific to TSP-$PM_{10}$, explaining nearly all TSP-$PM_{10}$, with negligible contributions to all other pollutants. This is consistent with Figure S1, in which TSP-$PM_{10}$ shows weak correlations with most other pollutants, suggesting a split between larger (TSP-$PM_{10}$) and finer ($PM_1$; $PM_{2.5}$-$PM_1$; and $PM_{10}$-$PM_{2.5}$) PM. Factor 2 is the primary driver of CO and contributes around 70% of each of BC, NO, and $NO_2$, with 25% of $PM_1$ and $PM_{10}$-$PM_{2.5}$. Uncertainty is largest for factor attribution for $PM_{10}$-$PM_{2.5}$, with Factors 1 and 2 competing to attribute its concentrations. These estimated factor attributions motivated targeted downstream analyses for the nearby coal terminal and traffic: coarse PM and BC are commonly associated with industrial dust sources, while CO and NO are characteristic of traffic-related combustion. Each GeomNMF bootstrap run completed in approximately 10 minutes on a standard laptop (2021 MacBook Pro, Apple M1 Pro, 16 GB RAM) for this nearly half-a-million observation dataset, confirming practical scalability.

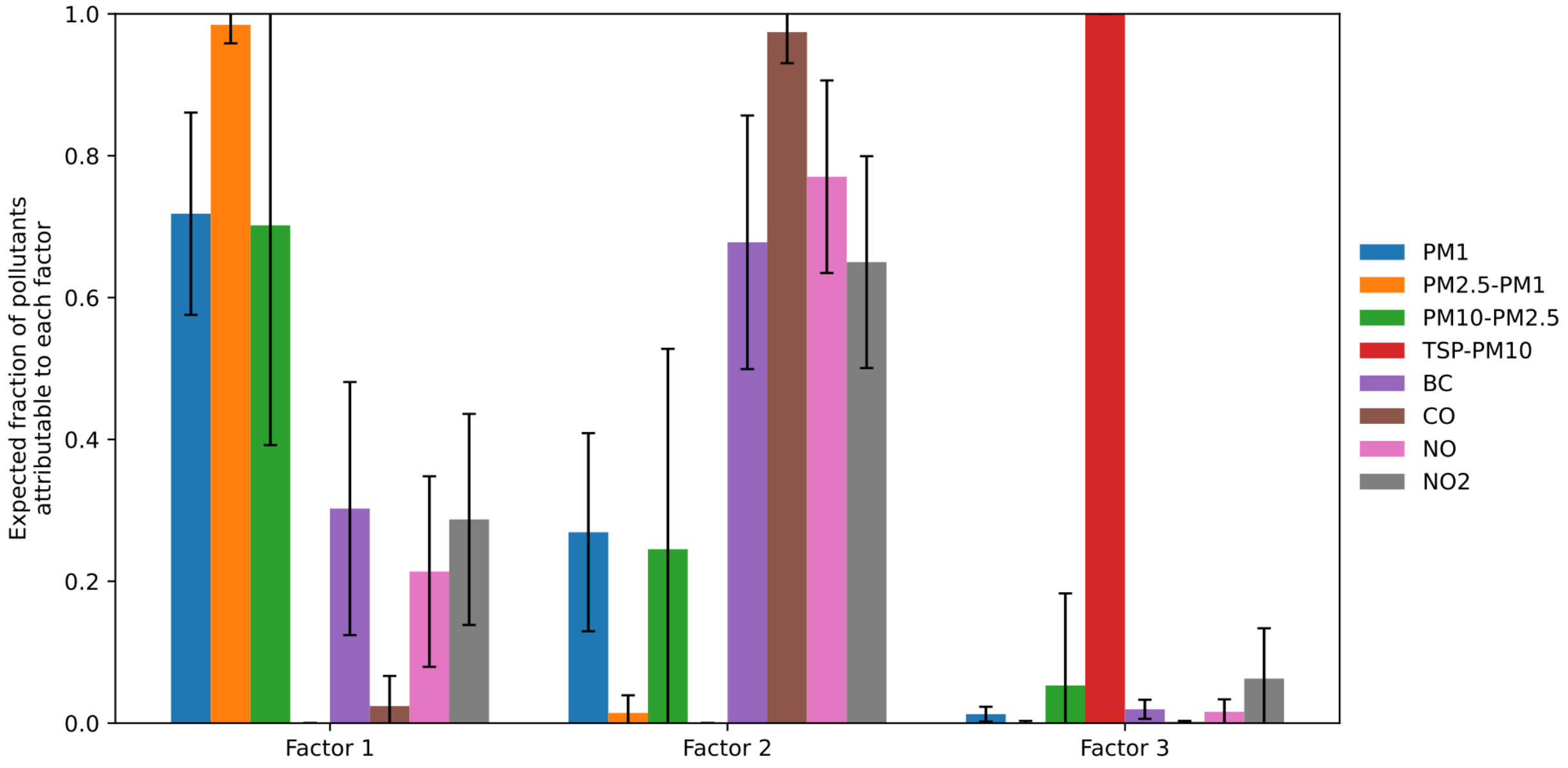

**Figure 3.** Bootstrap mean ± SE of the estimated factor attribution matrix $\Phi$ from GeomNMF ($K = 3$) across 100 resamples. Each bar represents the expected fraction of a given pollutant attributable to a factor, with attributions summing to 1 across factors for each pollutant. *Note.* PM = particulate matter; TSP = total suspended particles; BC = black carbon; CO = carbon monoxide; NO = nitric oxide; $NO_2$ = nitrogen dioxide.

We choose $K = 3$ because it yields a more coherent interpretation than larger models ($K \in \{4, 5\}$, see Figure S2, Supporting Information), which primarily split existing factors along pollutants with higher bootstrap uncertainty, such as BC, $PM_{10}$-$PM_{2.5}$, and NO, rather than uncovering physically meaningful signals. The $K = 3$ solution performs best on the rank stability metric (Table S2, Supporting Information), which we prioritize over coefficients of variation because reliable identification of leading factors has clearer policy implications. The roughly triangular shape of the sample convex hull in most pairwise projections (Figure S3, Supporting Information) provides additional visual support for $K = 3$. We also verify that the probabilistic separability condition holds for $K = 3$: for each factor $k \in \{1,2,3\}$, the row-normalized version of $\widetilde{W}_{ik}$ (denoted by $W^*_{ik}$) should place non-negligible probability mass near one. The empirical distributions of $W^*_{ik}$ are consistent with this requirement (Figure S4, Supporting Information), indicating that for each factor, a fraction of samples is dominated by that factor. Overall, $K = 3$ represents a principled, data-driven choice given our data dimension ($J = 8$ pollutants).

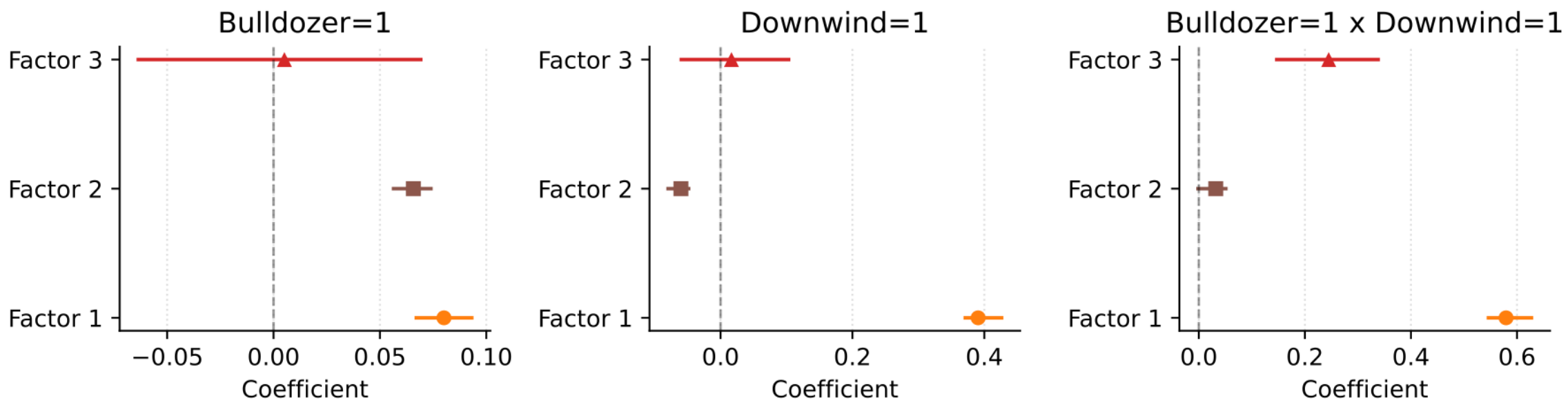

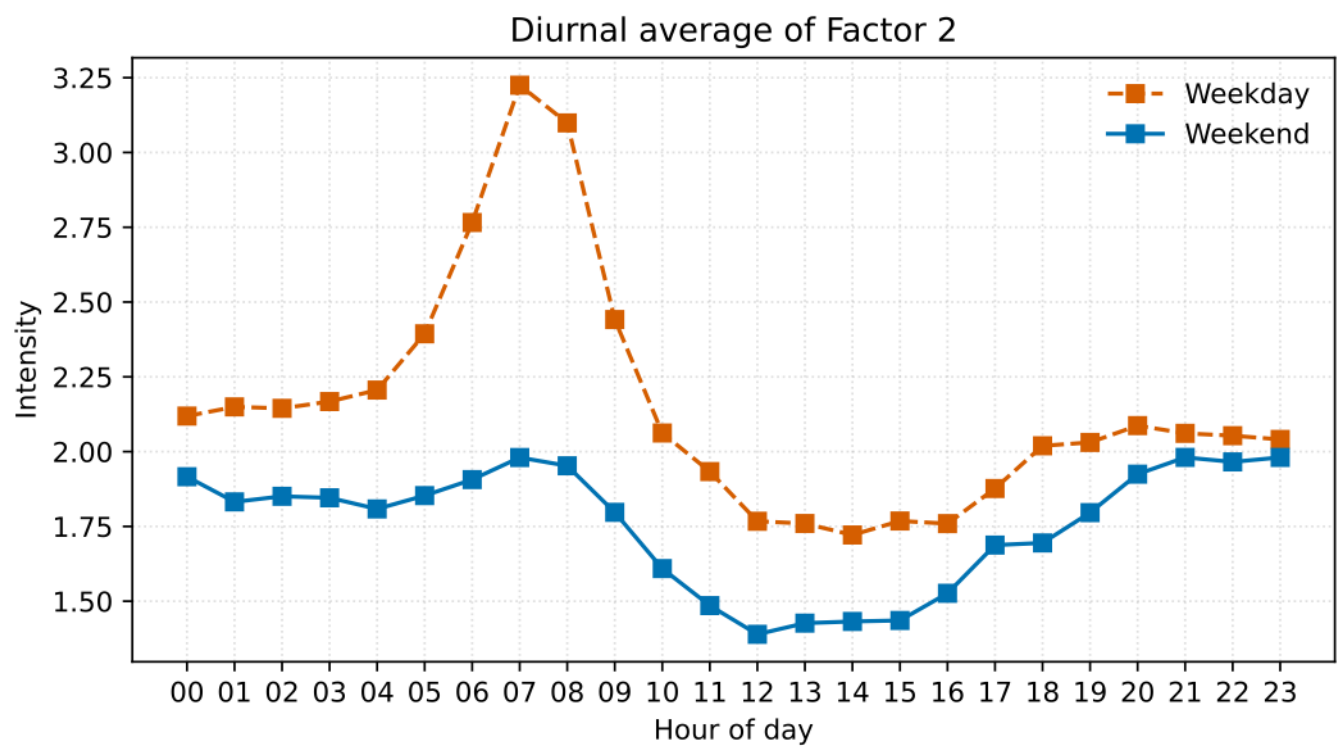

**Figure 4.** Top: Regression coefficients (bootstrap means with 95% CIs) for key covariates on estimated factor intensity, fit separately for each factor. Bulldozer=1 indicates minutes with visible bulldozer activity at the coal terminal. Downwind=1 indicates minutes when the wind direction was downwind of the coal terminal. Bottom: Diurnal averages of estimated Factor 2 intensity by weekday and weekend.

Figure 4 presents the downstream regression results from separate models fit to each estimated factor intensity, with visible bulldozer activity and downwind status as key predictors. Visible bulldozer activity occurred for an average of 460 minutes per day (32% of a day). Winds blew from the terminal toward Locations 1, 2, 5, and 8 for an average of 230 (16% of a day), 195 (14%), 83 (6%), and 128 (9%) minutes per day, respectively. The top panel reports bootstrap means and 95% confidence intervals (CIs) for the coefficients corresponding to Bulldozer=1, Downwind=1, and these main effects plus the interaction ("Bulldozer=1 × Downwind=1"). Factor 1 is positively associated with both Bulldozer=1 and Downwind=1, indicating a direct coal-terminal influence. Factor 3 is largely insensitive to the main effects but increases when bulldozer activity and downwind conditions coincide, suggesting an episodic contribution linked to terminal operations. This pattern reflects the dual requirement of active mechanical disturbance at the terminal and favorable wind to transport coarse particles to the monitors, consistent with the known physics that coarse particles are both mechanically lofted and gravitationally limited in transport range[41]. Factor 2 decreases when winds blow from the terminal, suggesting it is not terminal-related; instead, it exhibits clear weekday versus weekend contrasts in intensity and a pronounced weekday morning peak (~7 – 8 am; bottom panel, Figure 4), consistent with traffic-related emissions. This pattern mirrors the diurnal profiles of traffic-related pollutants that Factor 2 dominates (CO and $NO_2$[42,43]; Figure S5, Supporting Information). Independent traffic counts likewise confirm the weekday morning peak (Figure S6 in Supporting Information). The traffic record also shows an additional afternoon/evening peak which is not evident in the diurnal patterns of Factor 2, likely muted by planetary boundary layer dynamics that flatten late-day peaks[44]. Regression coefficient patterns for individual pollutants mirror those of their corresponding top-contributing factors (Figure S7, Supporting Information) and are consistent with pollutant-level regression results previously reported for this monitoring network[22]. Except for CO, most pollutants increase in concentration when bulldozer and downwind coincide, supporting the premise that the coal terminal activity is associated with elevated concentrations of these pollutants in Curtis Bay.

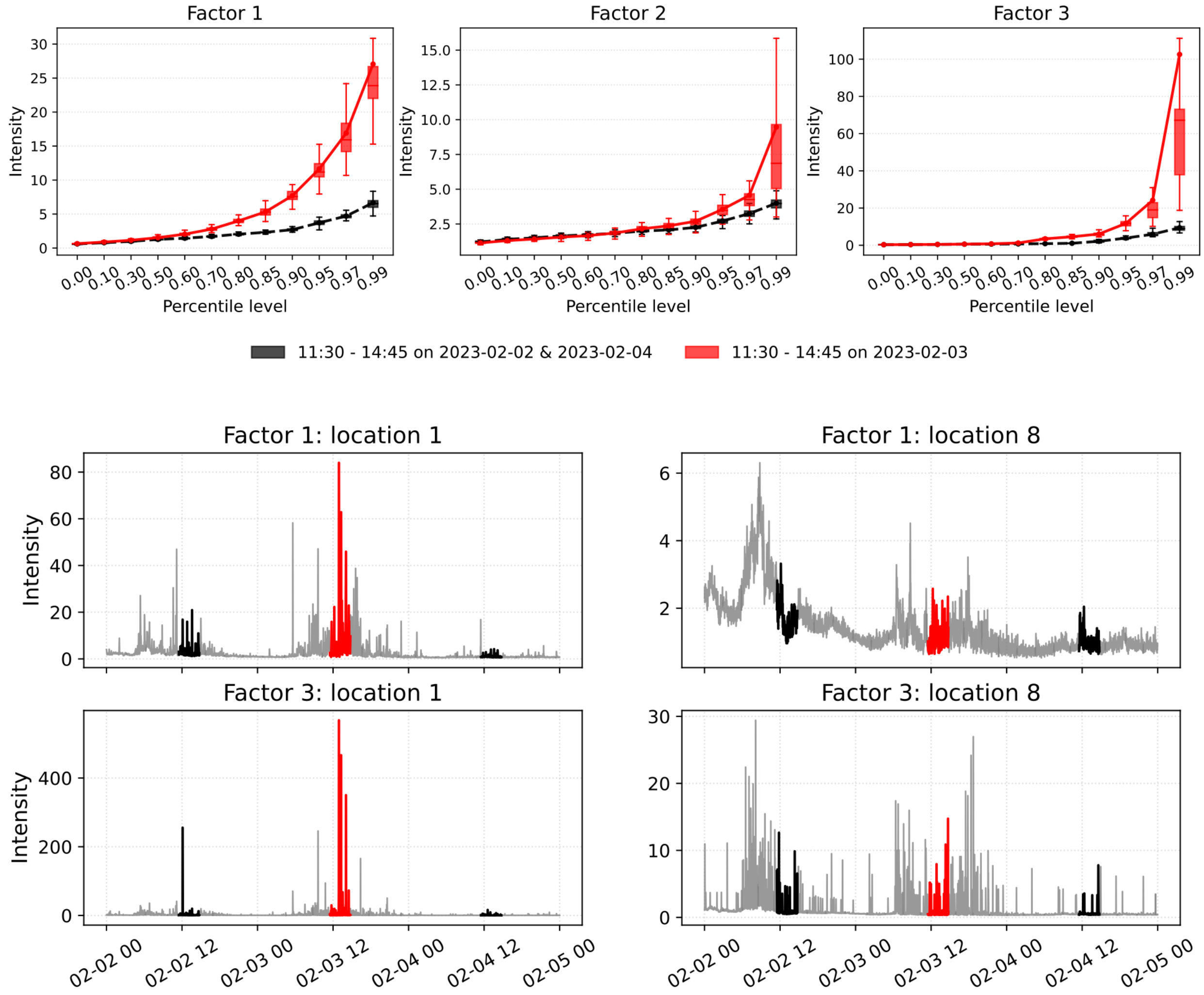

**Figure 5.** Comparison of air pollution levels during the coal-terminal incident (highlighted in red) versus matched hours (black) on the day before and the day after. Top: percentiles of estimated factor intensity across 100 bootstrap resamples; lines connect bootstrap means. Bottom: location-specific time series of estimated factor intensity for factors rejecting both the MW and KS tests at the 0.001 significance level.

Figure 5 presents the case study of the incident when a bulldozer was reported to be immobilized on a coal pile at the coal terminal on February 3, 2023 (11:30 am to 2:45 pm local time). This episode is atypical relative to most days in the study period and used to illustrate how the identified factors respond under an acute terminal event. Incident hours are compared against the same clock hours on the preceding (February 2) and following (February 4) days to control for diurnal and operational patterns, shown in red and black respectively.

The top panel of Figure 5 displays boxplots of bootstrap distributions of intensity percentiles across 100 resamples. Factor 1 shows a clear and substantial elevation beginning near the 70$^{th}$ percentile that becomes more pronounced at higher percentiles, whereas Factor 3 exhibits heavier upper tails (prominent after 95$^{th}$ percentile) during the incident with greater high-end variability. Factor 2, while elevated during the incident, has more uncertainty and overlap in two distributions. These findings echo Figure 4, where Factors 1 and 3 are more strongly related with terminal-related events than Factor 2.

These visual patterns are confirmed by the MW and KS tests. Factors 1 and 3 have significantly elevated intensities during the incident period under both tests ($p$-value $< 0.001$). Factor 2 shows no clear evidence for elevated intensity: its KS p-value is 0.016 and the MW p-value is 0.998, suggesting there may be sporadic high intensity during incident hours but no overall stochastic dominance. The bottom panel shows location-specific time series of estimated intensity for Factors 1 and 3 at the two monitoring sites that recorded data during this window (Locations 1 and 8). Location 1, which is 0.33 km from the centroid of the terminal, shows pronounced spikes during the incident relative to the matched hours, while the effects are attenuated at Location 8, which is 1.22 km away.

Pollutant-level analysis (Table S3, Supporting Information) corroborates these findings. Using pooled data from Locations 1 and 8, both the MW and KS tests reject the null hypothesis for $PM_{2.5}$-$PM_1$, $PM_{10}$-$PM_{2.5}$, and TSP-$PM_{10}$, with p-values < 0.001 in both tests. Testing separately by location, at Location 1, both tests reject the null hypothesis for BC in addition to $PM_{2.5}$-$PM_1$, $PM_{10}$-$PM_{2.5}$, and TSP-$PM_{10}$, with p-values ≤ 0.005. At Location 8, both tests find significant elevation only for $PM_{2.5}$-$PM_1$ at the same threshold. This spatial contrast is consistent with proximity-dependent impacts, as finer particles can travel farther from the terminal while coarser fractions are less likely to reach greater distances. Other pollutants do not reject the null in either test even at the 0.01 significance level. Taken together, these results suggest that the elevation of Factors 1 and 3 during the incident is driven primarily by mechanically generated dust.

Finally, we address community-raised questions about the frequency and duration of extreme coal dust events and how these patterns vary with distance from the coal terminal in Curtis Bay. We consider Factors 1 and 3 as indicative of the occurrence of mechanically generated dark dust associated with coal terminal activities. Table 2 shows that, at Location 1, the highest intensity events exceeding the 99$^{th}$ percentile of Factor 1 occur once every two hours and last about three consecutive minutes on average. During the study period, the longest such event lasted 114 minutes, and approximately 33 minutes per day were affected by these highest-intensity episodes. High intensity Factor 1 events occur more often per hour and have longer maximum durations at locations closer to the coal terminal, which is also reflected in the exceedance rates. Because thresholds are defined as percentiles, the nominal exceedance rates for the 90$^{th}$, 95$^{th}$, and 99$^{th}$ percentile thresholds are 10%, 5%, and 1%, respectively. At Location 1, the observed exceedance rates (17%, 9%, and 2%, respectively) exceed these nominal values whereas at locations farther from the terminal (Locations 2, 5, and 8, in order of increasing distance from the terminal centroid) the exceedance rates are below the nominal rates. As shown in Table 2, exceedance events are short-lived, with mean durations of 2.5 to 7.3 minutes. Across locations and thresholds, the percentage of events shorter than 5 minutes ranges from 76% (Location 8, 95$^{th}$ percentile) to 90% (Location 1, 99$^{th}$ percentile; not shown in the table). Minute-level resolution is therefore essential to capture this fine temporal structure, which coarser aggregation (e.g., hourly averaging) cannot resolve.

**Table 2.** Summary of Factor 1 intensity above percentile-based thresholds. *Exceedance events* are counts of continuous episodes (including single-minute episodes) during which intensity remains above the threshold. *Exceedance rate* is the percentage of one-minute samples above the threshold. *Duration* is the length of each exceedance event in minutes.

| Location | Threshold | Exceedance events (count) | Exceedance events /Hour (count) | Exceedance rate (%) | Mean duration per event (min) | Max duration per event (min) | Mean total duration /Day (min) |
|---|---|---|---|---|---|---|---|
| 1 | 90$^{th}$ | 5079 | 1.9 | 17.0 | 5.3 | 1824 | 244.7 |
| | 95$^{th}$ | 3434 | 1.3 | 9.3 | 4.3 | 875 | 133.5 |
| | 99$^{th}$ | 1369 | 0.5 | 2.3 | 2.6 | 114 | 32.7 |
| 2 | 90$^{th}$ | 1693 | 0.9 | 7.2 | 4.7 | 429 | 104.0 |
| | 95$^{th}$ | 871 | 0.5 | 3.3 | 4.1 | 212 | 47.2 |
| | 99$^{th}$ | 244 | 0.1 | 0.6 | 2.5 | 28 | 8.0 |
| 5 | 90$^{th}$ | 248 | 0.4 | 4.9 | 7.3 | 495 | 70.8 |
| | 95$^{th}$ | 110 | 0.2 | 1.3 | 4.2 | 54 | 18.2 |
| | 99$^{th}$ | 8 | 0.01 | 0.1 | 3.5 | 17 | 1.1 |
| 8 | 90$^{th}$ | 1238 | 0.5 | 5.8 | 6.9 | 429 | 84.0 |
| | 95$^{th}$ | 650 | 0.3 | 2.6 | 5.9 | 166 | 37.9 |
| | 99$^{th}$ | 89 | 0.04 | 0.2 | 3.3 | 31 | 2.8 |

Factor 3 shows similar patterns in exceedance events, rate, and duration, but with generally lower magnitudes (Table S4, Supporting Information). At Location 1, the highest-intensity Factor 3 events exceeding the 99th percentile occur roughly 0.8 times per hour, with a maximum event duration of 23 minutes and approximately 24 minutes per day affected during the study period. By contrast, Factor 2 does not exhibit such a prominent trend with distance to the coal terminal (Table S5, Supporting Information). The clear gradient in exceedance rates and event durations with distance from the terminal in Factors 1 and 3 demonstrates that our method resolves meaningful local spatial structure rather than reflecting spatially uniform regional background. Combined with the temporal dynamics during the bulldozer incident in Figure 5, these local patterns constitute actionable, quantitative evidence that is directly relevant to community air quality monitoring and policy.

We now turn to sensitivity analyses assessing the robustness of the main findings. Estimates of $\Phi$ are highly similar between GeomNMF and XRAY for the same number of factors $K$ (Figure S8 in Supporting Information). Least-squares NMF yields broadly similar patterns but with important discrepancies, especially for PM components and BC (Figure S9, Supporting Information). Moreover, the estimates change depending on whether an L2 penalty is used, with differences far exceeding bootstrap uncertainty. This sensitivity to penalty choice undermines the practical reliability of least-squares NMF, since the appropriate degree of regularization is rarely known in advance or can be well chosen from the data. Furthermore, the very small bootstrap SEs from least-squares NMF and correspondingly narrow CIs fail to reflect the substantial sensitivity of estimates to penalty choice, giving a misleading impression of precision.

Estimates of $\Phi$ from GeomNMF are robust to measurement uncertainty at the reported levels. Relative to Figure 3 and Table S1, Table S6 in Supporting Information shows that the mean estimates of $\Phi$ are nearly unchanged across datasets perturbed at reported uncertainty levels (see Sensitivity analysis for details), with the ranking of factor contributions preserved for each pollutant. Moreover, Monte Carlo SEs under this perturbation are comparable to the bootstrap SEs from the original data. Overall, this indicates that measurement uncertainty at the reported levels is of similar magnitude to sampling variability and has limited impact on our findings.

In the analysis excluding the two highest-uncertainty pollutants ($PM_{10}$-$PM_{2.5}$ and $NO_2$), the estimated $\Phi$ restricted to six pollutants stays nearly the same as the original estimates (Figure S10 in Supporting Information). Changes are within at most one bootstrap SE. This indicates that the identified factor attributions with three factors are not driven by the most uncertain pollutants.

Factor interpretation is robust to PM size fraction definitions. Regrouping PM variables as either $PM_1$ and TSP-$PM_1$, or $PM_{2.5}$ and TSP-$PM_{2.5}$, yields qualitatively identical conclusions (Figure S11, Supporting Information). This confirms that our inference is not an artifact of how PM size fractions were partitioned, and that the key signals are carried by the coarser fractions where mechanical emissions dominate. The estimates of $\Phi$ remain stable under alternative normalization percentiles ($90^{th}$ and $95^{th}$; Figure S12 in Supporting Information), with bootstrap uncertainties increasing slightly at the higher percentiles for some elements. When we refit the model to a reduced dataset ($n = 451{,}721$) that excludes the 30 largest observations for each pollutant, removing 0.05% of the original data, the estimates of $\Phi$ are close to those from the full data (Figure S13 in Supporting Information). In particular, TSP-$PM_{10}$ is again attributed mainly to Factor 3, indicating that this attribution is not driven by a few extreme TSP-$PM_{10}$ values. The factor attribution persists even after the maximum of TSP-$PM_{10}$ drops from 17503.5 to 9234.1 $\mu g/m^3$ (a 47% reduction).

## Discussion

This study provides, to the best of our knowledge, the first application of a scalable, reliable geometric NMF approach to source apportionment of hyperlocal, high temporal resolution multi-pollutant air sensor network data. This approach allowed investigation of hypotheses related to specific industrial facilities, events, and activities of concern to local community members that have been difficult to disentangle within the context of cumulative impacts and local air pollution burdens.

The geometric NMF analysis revealed strong evidence of factors related to traffic and the coal terminal in Curtis Bay. Among these, Factors 1 and 3 emerged as strongly associated with terminal operations, contributing to PM across size fractions and BC. This pollutant profile has important implications for the health of nearby residents. Fine PM ($PM_1$ and $PM_{2.5}$) penetrate deep into the respiratory tract and bloodstream, exacerbating cardiovascular and respiratory conditions, while acute exposure to $PM_{10}$ has been associated with increased hospital admissions for asthma, bronchitis, and chronic obstructive pulmonary disease and heightened risks of stroke and cardiac events[45,46]. In contrast, the burden of the TSP- $PM_{10}$ fraction on nearby residents operates substantially through non-inhalation pathways. Particles in this size range settle rapidly onto soil, water, gardens, and outdoor surfaces, where they contaminate the residential environment and re-enter the body through incidental ingestion (e.g., hand-to-mouth contact in young children). Beyond these exposures, the coarser fraction contributes to welfare effects that shape residents' daily environment, including soiling of homes and property and locally reduced visibility from fugitive dust, which is central to the disproportionate environmental concerns of the Curtis Bay community. Acute and long-term exposure to BC has been linked to cardiovascular and respiratory morbidity and mortality[47,48] and to adverse fetal, pediatric, and maternal health outcomes[49]. When these pollutants co-occur from a common pollution origin, their individual risks may compound. Future studies linking factor intensity estimates by geometric NMF to health outcomes in longitudinal, repeated-measures epidemiologic studies could quantify the joint health impacts of these co-occurring pollutants. Such studies could examine biologically relevant outcomes including respiratory and cardiovascular disease, health-related quality of life, and fetal and pediatric health.

The present study has multiple strengths. First, the geometric source apportionment framework addresses key limitations of existing methods in terms of both statistical rigor and practical accessibility. Our framework targets the interpretable factor attributions directly, avoiding scale ambiguity and yielding better-calibrated uncertainty quantification that supports principled inference. Unlike PMF and other traditional methods, it scales to large, high-frequency datasets and supports straightforward application to non-speciated multi-pollutant data without requiring proprietary software. Its open Python implementation lowers the barrier to adoption in community air quality studies and makes the analysis fully reproducible. Second, as a community-driven research output, community members from Curtis Bay and South Baltimore, through CCBA and SBCLT, guided the research questions investigated, supported deployment of multi-pollutant air sensors, and contributed to the interpretation of the data analysis and its outputs. This included the evaluation of the coherence of the present study's findings with those of prior studies that confirmed the presence of coal dust within settled dark dust collected in Curtis Bay[23] and that demonstrated increased PM and BC air pollution burden related to the coal terminal in Curtis Bay[22]. The participatory approach of these studies improves the relevance of the scientific findings to answer longstanding community questions about complex exposure and health burdens as well as to inform policy and regulatory decision-making that can mitigate emitter-specific burdens of air pollution exposure.

This study has several limitations. Because air sensor data are non-speciated, certain source types, including regional background and secondary aerosol, that are conventionally distinguished through speciated measurements cannot be separately identified. This is a constraint of the measurement rather than of the apportionment framework: without species such as sulfate and nitrate, those source types cannot be resolved by any receptor method. It is therefore reasonable to regard those types as subsumed within the estimated factors rather than absent. For instance, Factor 1 accounts for a large fraction of the fine PM. Regional background shares this fine-mode signature but has no separate chemical tracer in the data. Some regional contribution is therefore likely carried within Factor 1, indistinguishable from the coal terminal related signal. While this apportionment does not resolve the full set of source types recoverable with speciated data, our results remain robust and interpretable through downstream analyses grounded in community concerns and reported events. Regarding downstream regression analyses, we investigated associations of the inferred factors with downwind direction and bulldozer activity at 1-minute time resolution, without investigating potential time-lag effects. Other visible coal terminal activity patterns are also likely to influence community air pollution burden, including train transport and unloading[50], rail track cleaning[24], and ship loading. Future studies should extend our time-resolved analysis of bulldozer activity to these patterns. Finally, examining exposure-response associations at time lags informed by the fate, transport, and dispersion dynamics of varying particle size distributions would further strengthen the epidemiological relevance of this approach. From the community perspective, this analysis identified two predominant drivers of air pollution burden in Curtis Bay: traffic and the coal terminal. However, Curtis Bay residents face a much broader environmental burden from dozens of overlapping contributors, including medical waste incineration, hazardous and non-hazardous waste landfills, incinerator ash disposal, chemical plants, fossil fuel storage and recycling operations, wastewater treatment, and animal rendering facilities. Future work should examine their cumulative and potentially synergistic effects on the community air pollution burden.

This analysis offers a transferable workflow for source apportionment in fenceline communities. GeomNMF is a practical and scalable tool for quantifying and separating drivers of air pollution burden, supporting evidence-based policy to reduce exposure and improve health. Beyond Curtis Bay, this workflow applies to communities bordering coal handling, transport, and export infrastructure such as Newport News, Virginia. More broadly, it extends to other locally-unwanted land uses including landfills, wastewater treatment plants, and industrial livestock operations, where attributing proximal offsite pollution burdens to routine operations remains a persistent challenge.

ASSOCIATED CONTENT

Supporting Information: Additional details for data and analyses and complementary figures and tables supporting the manuscript's main results (PDF).

AUTHOR INFORMATION

**Corresponding Authors**

*bora.jin@mines.edu

*cheaney1@jhu.edu

*abhidatta@jhu.edu

**Author Contributions**

‡These authors contributed equally as senior authors.

**Present Addresses**

David McClosky's present address is Rhizome Data, 3619 Georgia Avenue Northwest, Suite 502, Washington, DC, 20010, USA.

Laura E. Schmidt's present address is Data Science Institute, University of Chicago, 5460 S University Ave, Chicago, IL, 60615, USA.

**Funding Sources**

This work is partially supported by National Institute of Environmental Health Sciences (NIEHS) grant R01ES033739 and NIEHS P30 Center for Community Health: Addressing Regional Maryland Environmental Determinants of Disease (CHARMED) [grant no. P30ES032756]. BDS, MAA, LES, and CDH were supported by the Johns Hopkins Community Science and Innovation for Environmental Justice (CSI EJ) Initiative. MAA and CDH were supported by NIEHS grant no. P30ES032756. CDH was supported by the National Institute for Occupational Safety and Health (NIOSH) Education and Research Center [grant no. T42OH0008428]. LND was supported by NIEHS Training Program in Environmental Health Sciences [grant no. T32ES007141]. NJS was supported by the UC Davis Environmental Health Science Center Core [grant no. P30ES023513]. RRD was supported by the University of Maryland Challenge Grant and the Maryland Department of the Environment. RRD and the CCBA were supported by the EPA American Rescue Plan Enhanced Air Quality Monitoring for Communities (EPA-OAR-OAQPS-22-01) [grant no. U00P4600952].

ACKNOWLEDGMENT

We thank the Curtis Bay residents, local business owners, and faith-based organizations who trusted our team, contributed their lived experiences and knowledge, and hosted air quality monitors and/or trail cameras. We thank Carlos C. Sanchez-Gonzales, David Jones, and Angela Shaneyfelt for their contributions to and support of this work. We thank the students in the environmental justice class at Benjamin Franklin High School for their engagement and enthusiasm. We thank the MDE Air and Radiation Administration for their technical support and collaboration. We thank Anna Hodshire and Eben Cross from QuantAQ and Troy Cados, Julien Caubel, and Nick Wong from Distributed Sensing Technologies for their scientific guidance and technical support.

CONFLICTS OF INTEREST

MAA serves as an unpaid member of the board of directors of SBCLT, starting on June 24, 2024. The other authors declare that they have no known competing financial interests or personal relationships that could have appeared to influence the work reported in this manuscript.

# Supporting Information for
# Source apportionment of air pollution burden using geometric non-negative matrix factorization and high-throughput multi-pollutant air sensor data in Curtis Bay, Baltimore, USA

## S1. Materials and Methods

### Data

MODULAIR units showed strong agreement, per EPA Non-regulatory Supplemental and Informational Monitoring (NSIM) criteria, when compared with regulatory-grade measurements of $PM_{2.5}$ and $PM_{10}$[1–3] and of CO and $NO_2$[3]. Fine-particle measurements ($PM_1$ and $PM_{2.5}$) may be more reliable than measurements of larger particle fractions because aspiration efficiencies decrease as particle size increases[1]. NSIM does not specify targets for TSP or NO. DSTech ObservAir monitors measuring BC were also collocated with a regulatory aethalometer (Model AE33, Magee Scientific) at an Maryland Department of the Environment (MDE) monitoring site, and prior work has reported high agreement[4]. Based on this validation, no further adjustments were applied to the PM and gas measurements.

To study association of estimated factor intensities with traffic, data on traffic count and classification were collected using the Houston Radar Armadillo Tracker (Houston Radar LLC, Sugarland, TX) on Pennington Avenue and Curtis Avenue, two parallel routes in Curtis Bay. For each detected vehicle, the device records time and date of the event and vehicle class (small, medium or large vehicle). Per manufacturer specifications, the accuracy of vehicle counts is up to 97% for the local road configurations for typical free flowing traffic, with potential decreased accuracy for stop and go traffic[5]. Classifications were based on vehicle length with a length of < 4 m counted as small vehicles, between 4 and 7 m as medium vehicle, and > 7 m as large vehicles. Data were collected periodically between February 19, 2025, and November 5, 2025, totaling 134.8 days (3,234.2 hours) of observation. Although the time periods of the multi-pollutant and traffic datasets do not align, we expect the diurnal traffic pattern in the area to be stable over time. Given the proximity of the multi-pollutant air sensor locations and the traffic counters, the traffic data should reliably represent general diurnal trends in the study region.

### Methodological details

We adopt a scalability heuristic in the GeomNMF algorithm to handle very large datasets such as ours. For finding the maximum-volume $K$-subset of extreme points on the sample convex hull with $N$ corners, an exhaustive search over all $N$-choose-$K$ possibilities is computationally prohibitive. We therefore prune the $N$ hull vertices to retain a well-separated $N_0$ candidates ($N_0 \ll N$) through k-means clustering, then select $K$ representatives from those $N_0$ candidates that maximize the polytope volume. Balancing accuracy and cost, we set the pruned vertex set to have $N_0 = 40 \times K$, based on pilot experiments.

**Factor interpretation**

To quantify how frequently residents are exposed to air quality extremes attributable to local pollution origins, we summarize estimated factor intensities above high-value thresholds. Because factor intensities are highly right-skewed, we define thresholds using the factor-specific 90$^{th}$, 95$^{th}$, and 99$^{th}$ percentiles. We define an *exceedance event* as a continuous period during which the factor intensity remains above a given threshold. Such events may have different durations, and the *exceedance event count* is the number of such periods. *Exceedance events per hour* is computed as the event count divided by the total number of one-minute measurements, multiplied by 60. *Mean* and *maximum duration* refer to the mean and maximum lengths (in minutes) of exceedance events. In contrast, the *exceedance rate* is the percentage of one-minute measurements above the threshold. *Total* d*uration per day* is calculated as the exceedance rate multiplied by 60 × 24, representing the expected number of exceedance minutes per 24 hours. To capture spatial patterns, all summaries are computed separately for each location.

**Sensitivity analysis**

We attempted to fit EPA PMF 5.0 for comparison, as PMF incorporates measurement uncertainty through error-weighted fitting. Following prior guidance that EPA PMF can handle up to ~500,000 (sample × pollutant) combinations[6], we initially selected 60,000 samples with 8 pollutants (480,000 combinations) from the full dataset of 451,946 samples (3,615,568 combinations). However, EPA PMF returned a "system out of memory" error on an Intel Core machine with 8 GB RAM, preventing model fitting at that scale. The software ran only after reducing the sample size (to half or smaller), but base runs frequently failed to converge, even when we increased the number of runs beyond the recommended 20. When nonconvergence occurs, a common recommendation is to inflate uncertainty; we therefore tested tripling uncertainty for several pollutants, but this did not resolve the issue. Moreover, inflating uncertainty is not scientifically justified as a general remedy in our application. Nonconvergence difficulties with PMF have been reported elsewhere[7] as well. ESAT is a Python implementation optimizing the same weighted least-squares objective as PMF but using a different algorithm. It is compatible with macOS unlike EPA PMF 5.0. Despite improved computational efficiency, ESAT also failed to fit the full dataset on a 2021 MacBook Pro (Apple M1 Pro, 16 GB RAM) using Python 3.11. Although ESAT produced convergent results on subsampled datasets, subsetting introduces fundamental challenges including result aggregation across subsets, potential factor mismatch, and the emergence of subset-specific spurious factors, that can obscure holistic interpretation, and we did not proceed further.

## S2. Figures and tables

## <u>Data</u>

## Figure S1.

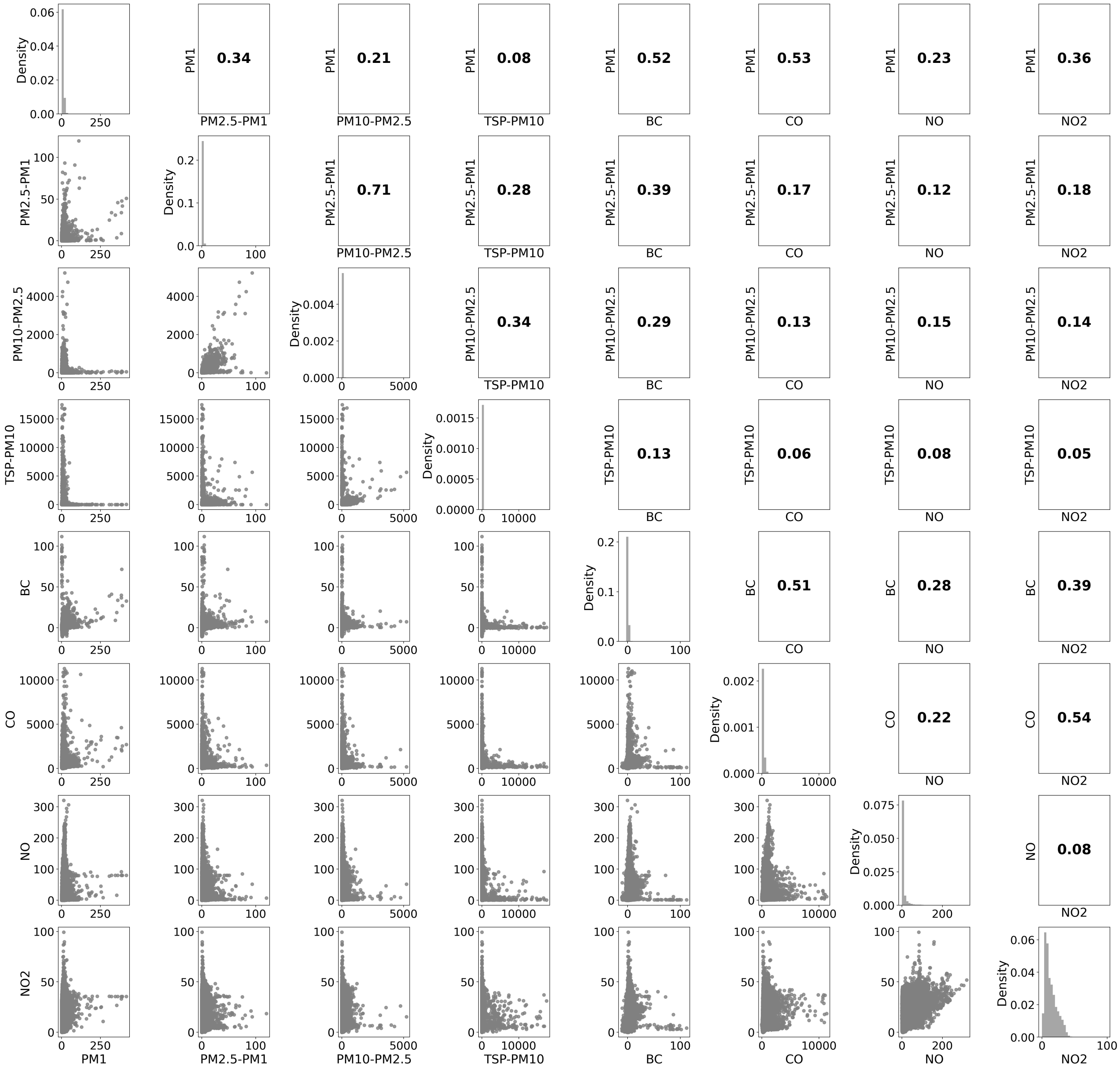

Matrix of air pollutant relationships for Curtis Bay: lower triangle shows pairwise scatter plots, the diagonal displays each pollutant's normalized histogram, and the upper triangle reports Spearman correlation coefficients. *Note.* PM = particulate matter; TSP = total suspended particles; BC = black carbon; CO = carbon monoxide; NO = nitric oxide; $NO_2$ = nitrogen dioxide.

In Figure S1, the skewness is evident from the marginal densities. The strongest correlation is observed between $PM_{2.5}$-$PM_1$ and $PM_{10}$-$PM_{2.5}$ (0.7). Moderate correlations (0.3 – 0.5) occur between several pairs of pollutants, e.g., adjacent PM size bins; $PM_1$ with gases; BC with all pollutants except TSP-$PM_{10}$; and CO with $NO_2$.

## Factor attribution results

**Table S1.** Bootstrap mean (bootstrap standard error; SE) of the estimated $\Phi$ by GeomNMF ($K = 3$)

| | **$PM_1$** | **$PM_{2.5}$-$PM_1$** | **$PM_{10}$-$PM_{2.5}$** | **TSP-$PM_{10}$** | **BC** | **CO** | **NO** | **$NO_2$** |
|---|---|---|---|---|---|---|---|---|
| Factor 1 | 0.718 (0.143) | 0.985 (0.026) | 0.702 (0.310) | 0 (<0.001) | 0.303 (0.178) | 0.024 (0.043) | 0.214 (0.134) | 0.287 (0.149) |
| Factor 2 | 0.269 (0.140) | 0.014 (0.025) | 0.245 (0.282) | 0 (<0.001) | 0.678 (0.179) | 0.974 (0.044) | 0.770 (0.136) | 0.650 (0.149) |
| Factor 3 | 0.013 (0.010) | 0.001 (0.002) | 0.053 (0.130) | 1 (<0.001) | 0.020 (0.014) | 0.002 (0.001) | 0.016 (0.018) | 0.063 (0.071) |

## Model checks

### Figure S2.

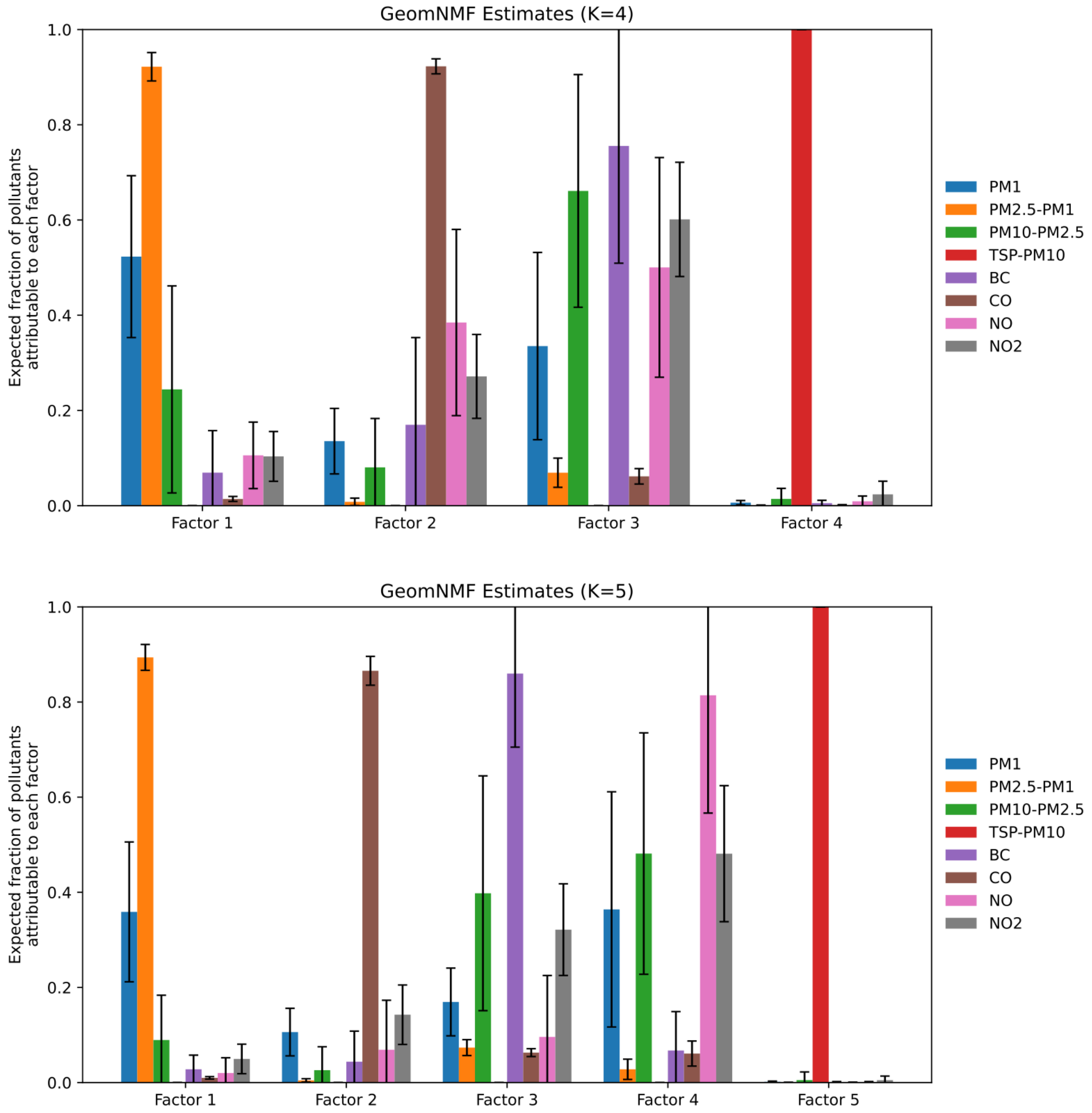

Bootstrap mean ± SE of the factor attribution matrix Φ estimated by GeomNMF, computed from 100 resamples for $K \in \{4,5\}$. The $K = 5$ solution appears to split Factor 3 from the $K = 4$ model into two factors (Factors 3 and 4).

**Table S2.** Diagnostics for Φ stability across $K$. CVar is the mean of cellwise coefficient of variations (bootstrap SE/bootstrap mean); lower values indicate more stable estimates. R is the mean of pollutant-specific rank stability; higher values indicate more consistent identification of the top contributor.

| | **CVar** | **R** |
|---|---|---|
| $\boldsymbol{K = 3}$ | 0.056 | 0.923 |
| $\boldsymbol{K = 4}$ | 0.050 | 0.859 |
| $\boldsymbol{K = 5}$ | 0.056 | 0.833 |

**Figure S3.**

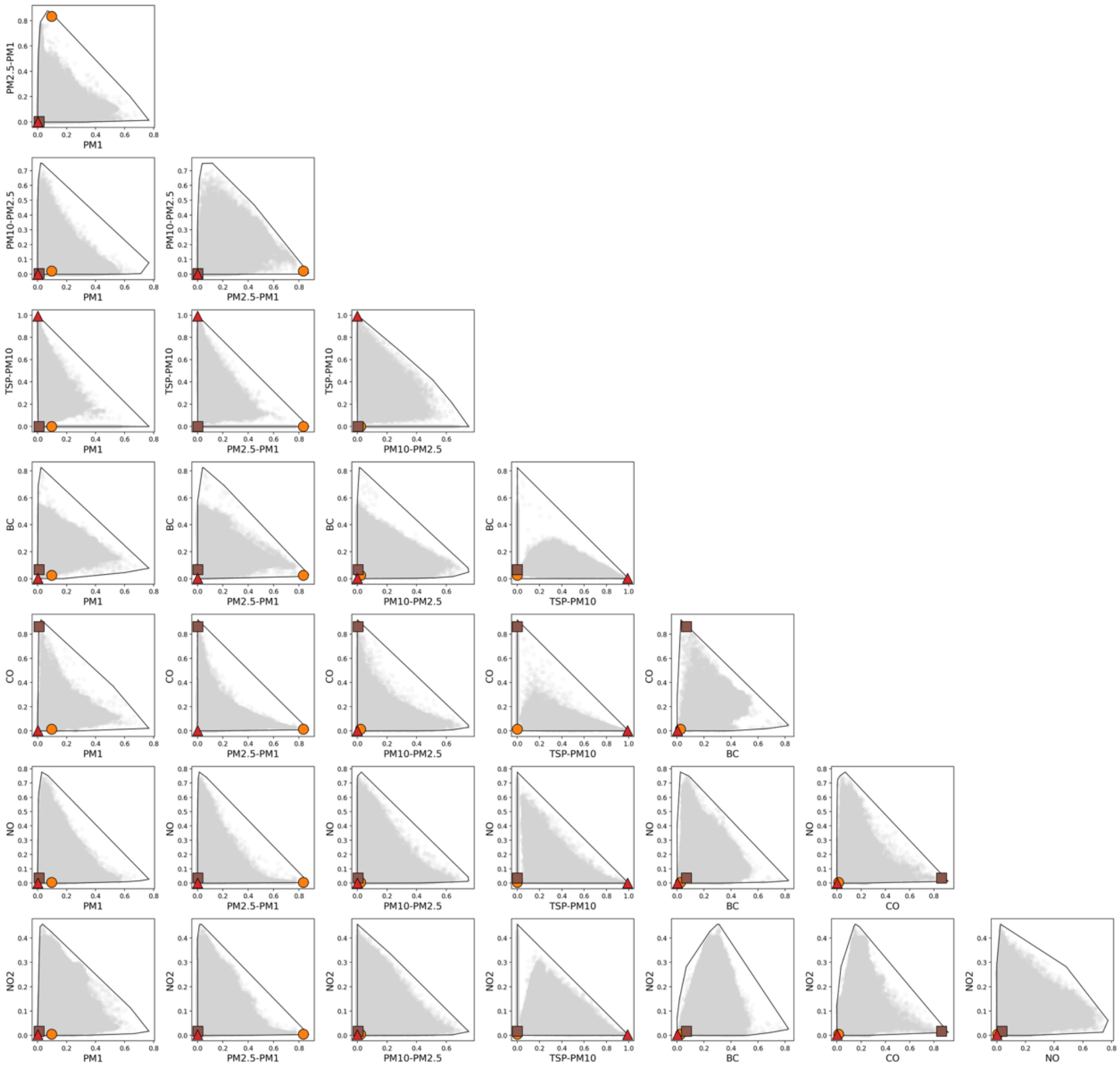

Boundary of sample convex hull (black lines) of the row-stochastic standardized pollutant concentrations $Y^*$, together with the estimated factor profile $\widehat{H}^*$ for $K = 3$ factors (orange circle, brown square, and red triangle).

**Figure S4.**

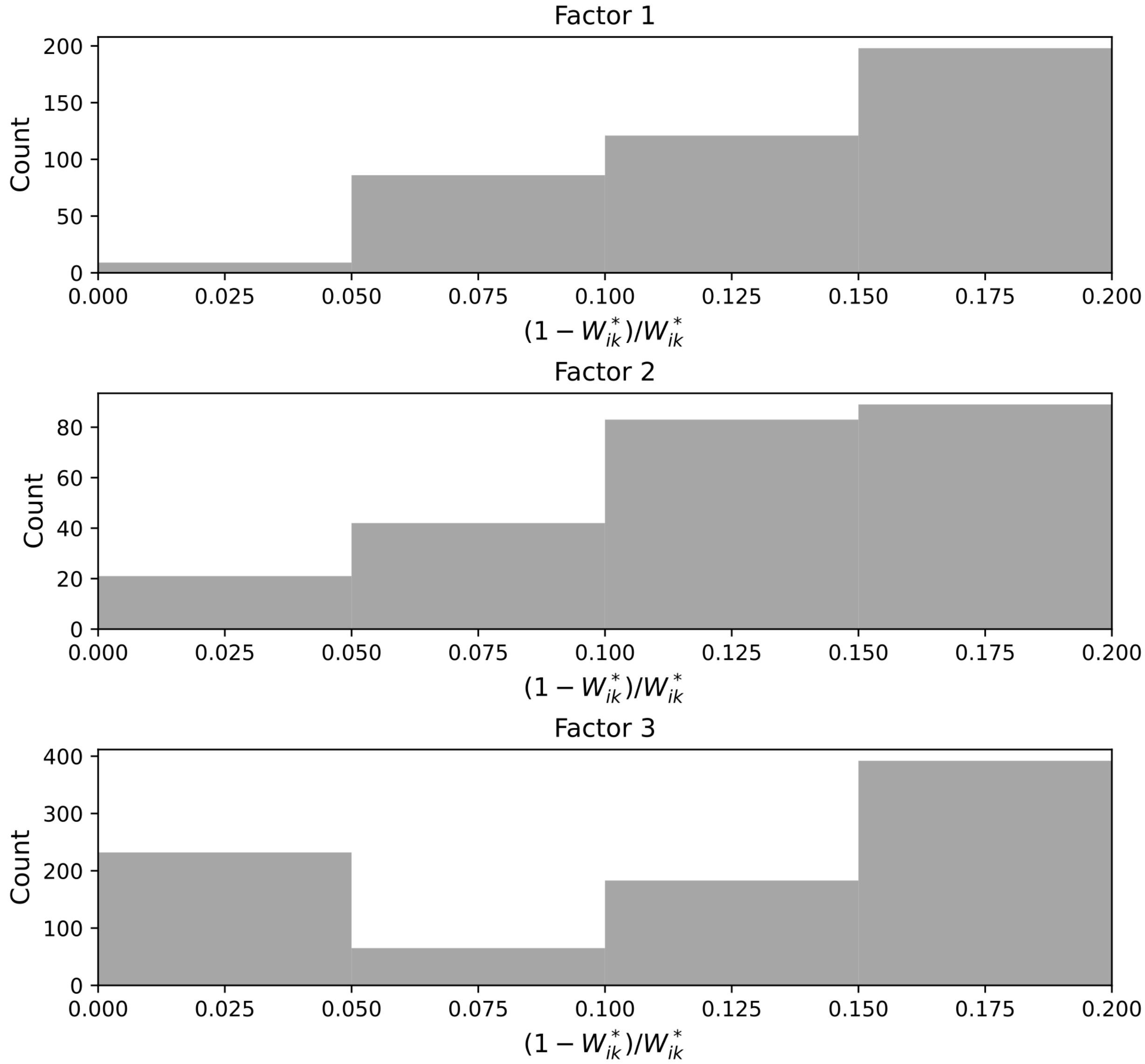

Assessing probabilistic separability: the ratio $(1 - W_{ik}^*)/W_{ik}^*$ should place nontrivial mass near 0. While it may diverge to infinity, here we restrict attention to [0, 0.2]. The matrix $W^*$ is row-normalized version of $\widetilde{W}$.

## Factor interpretation

**Figure S5.**

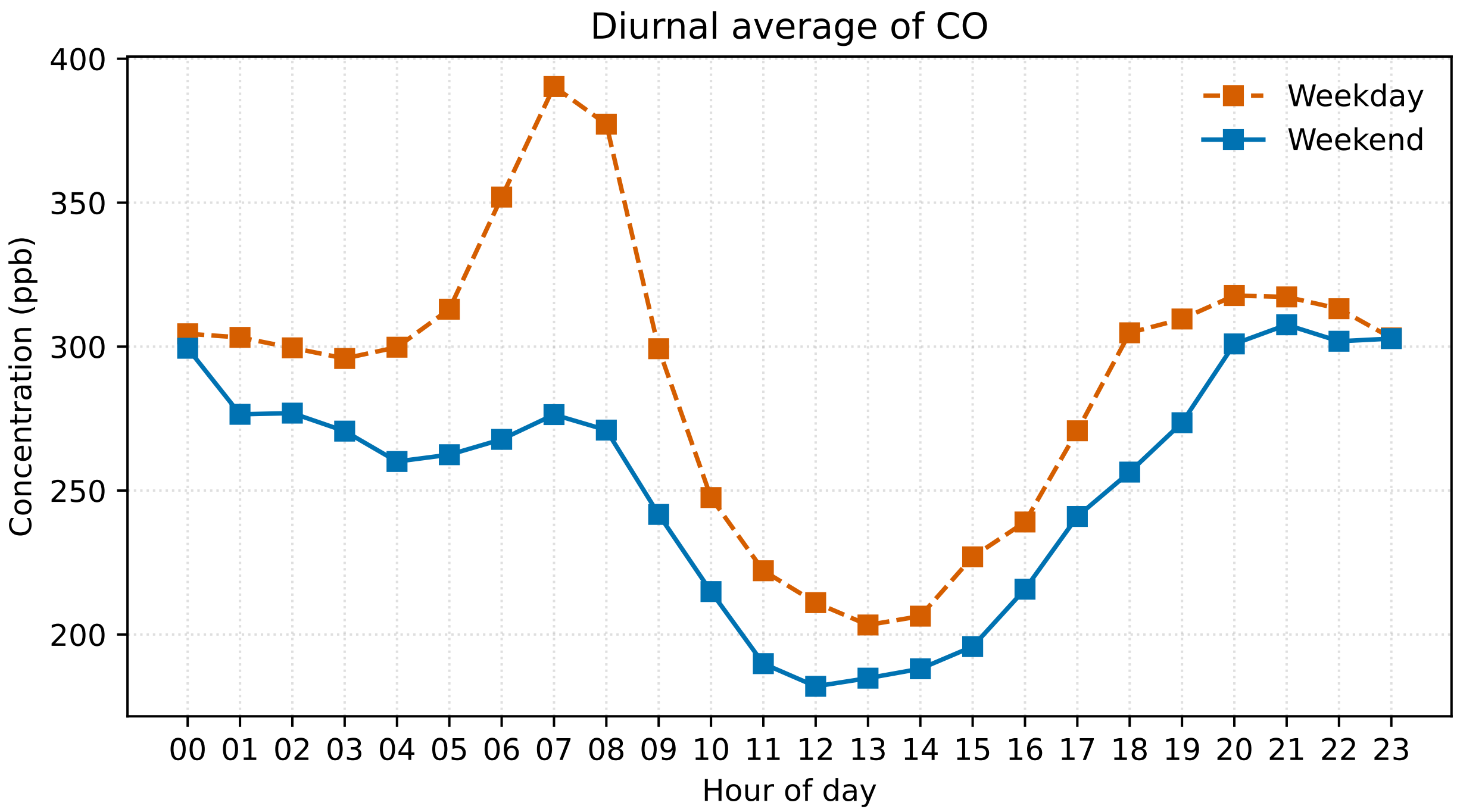

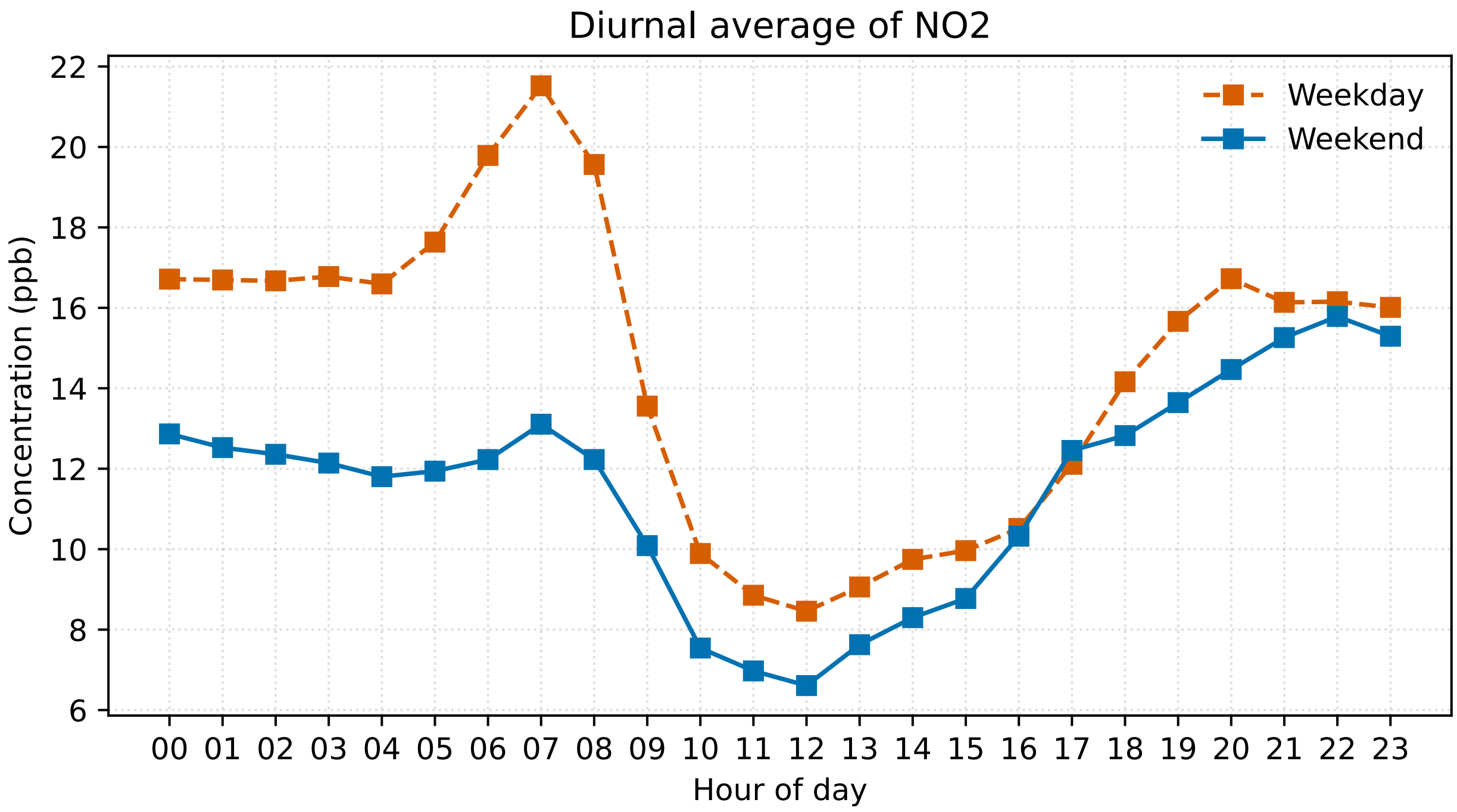

Diurnal averages of CO (top) and $NO_2$ (bottom) separately for weekdays and weekends. Factor 2 attributes 97% of CO and 65% of $NO_2$.

**Figure S6.**

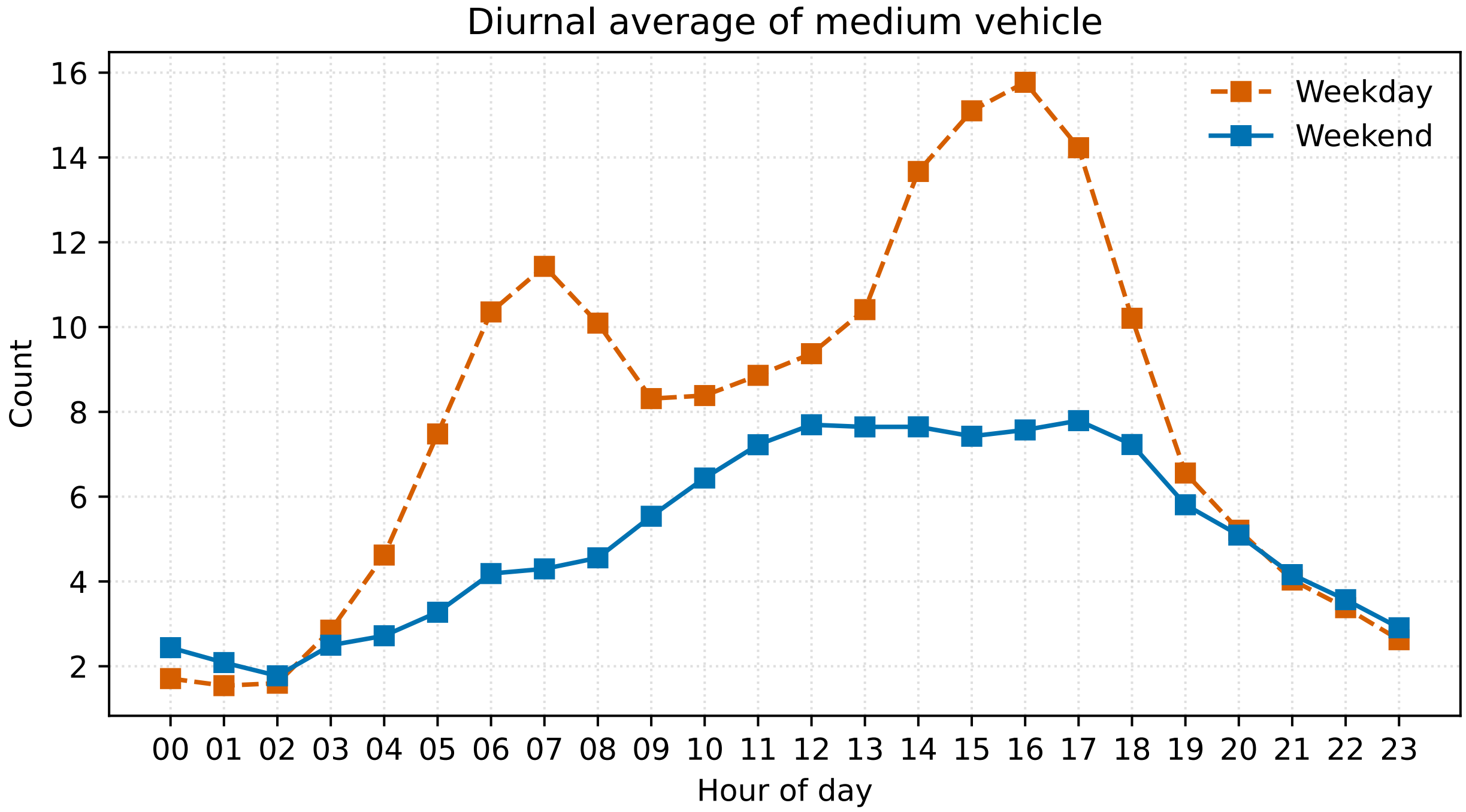

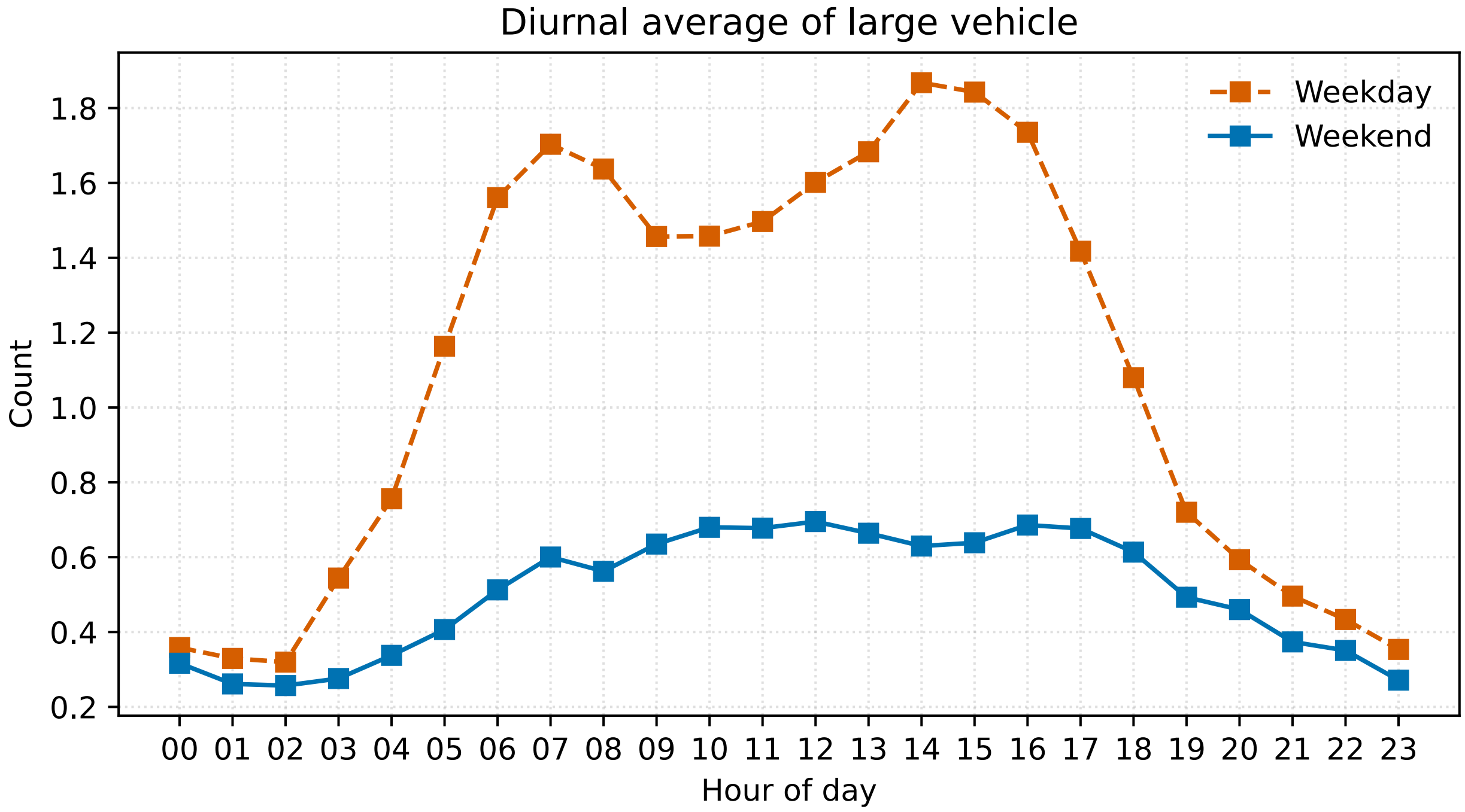

Diurnal averages of traffic counts by vehicle size, split by weekdays and weekends. "Medium" includes sedans and minivans; "large" includes delivery vans, buses, and dump trucks.

**Figure S7.**

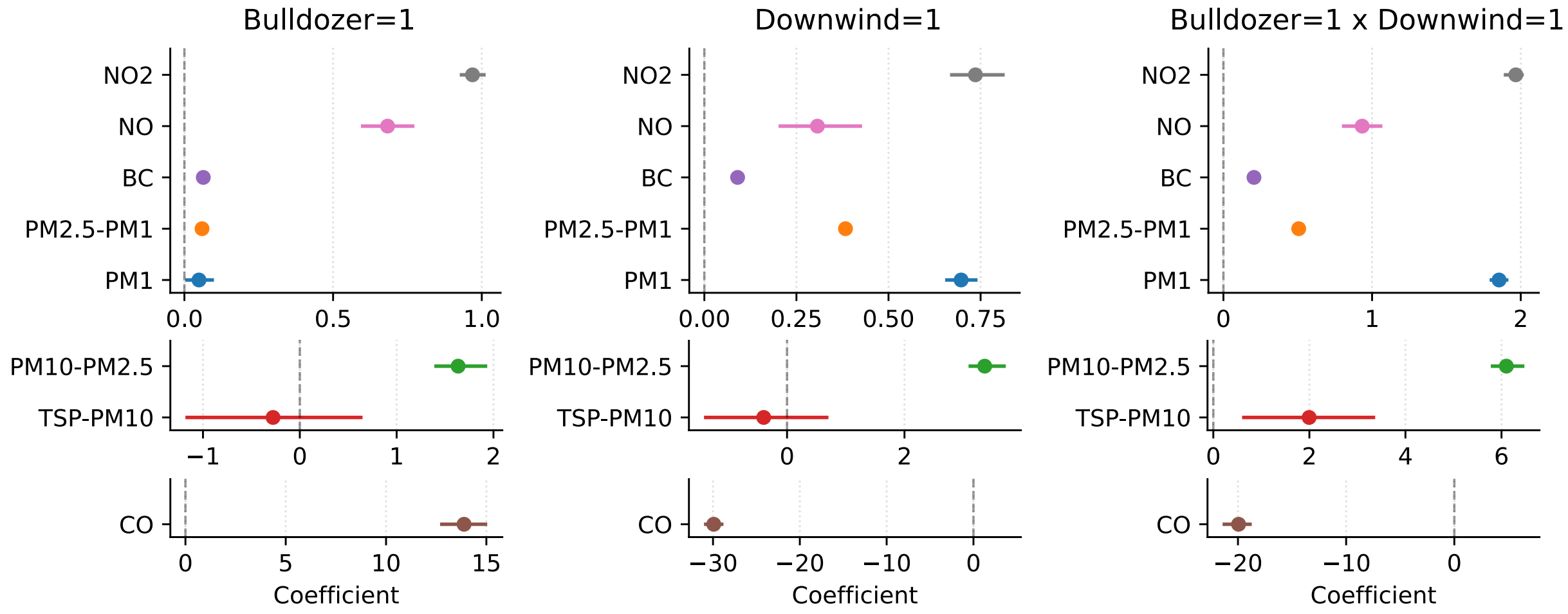

Regression coefficients (bootstrap means with 95% CIs) for key covariates on individual air pollutants. Because some coefficients are much larger in magnitude, results for $PM_{10}$-$PM_{2.5}$, TSP-$PM_{10}$, and CO are plotted separately.

**Table S3.** Summary statistics for air pollutants of interest at Locations 1 and 8 and pooled across both sites during the coal-terminal incident versus matched hours on the preceding and following days. We also report p-values from the MW and KS tests.

| ***Pooled*** | **Incident** | **Count** | **Mean (Std)** | **Median** | **95th percentile** | **Max** | **p-value (MW)** | **p-value (KS)** |
|---|---|---|---|---|---|---|---|---|
| $PM_{2.5} - PM1$ ($\mu g/m^3$) | 1 | 392 | 2.533 (5.012) | 0.782 | 10.633 | 61.349 | <0.001 | <0.001 |
| | 0 | 784 | 0.702 (1.205) | 0.439 | 1.916 | 17.984 | | |
| $PM_{10} - PM_{2.5}$ ($\mu g/m^3$) | 1 | 392 | 68.048 (262.682) | 22.931 | 176.825 | 3184.519 | <0.001 | <0.001 |
| | 0 | 784 | 13.987 (31.422) | 4.982 | 45.446 | 451.175 | | |
| $TSP - PM_{10}$ ($\mu g/m^3$) | 1 | 392 | 70.658 (532.121) | 0 | 145.961 | 7392.264 | <0.001 | <0.001 |
| | 0 | 784 | 9.659 (126.803) | 0 | 36.426 | 3509.289 | | |
| BC ($\mu g/m^3$) | 1 | 392 | 0.472 (0.708) | 0.352 | 1.819 | 3.562 | 1 | 0.170 |
| | 0 | 784 | 0.521 (0.458) | 0.487 | 1.336 | 2.683 | | |

| ***Location 1*** | **Incident** | **Count** | **Mean (Std)** | **Median** | **95th percentile** | **Max** | **p-value (MW)** | **p-value (KS)** |
|---|---|---|---|---|---|---|---|---|
| $PM_{2.5} - PM_1$ ($\mu g/m^3$) | 1 | 196 | 4.6 (6.5) | 2.4 | 15.4 | 61.3 | <0.001 | <0.001 |
| | 0 | 392 | 1.0 (1.6) | 0.5 | 2.7 | 18.0 | | |
| $PM_{10} - PM_{2.5}$ ($\mu g/m^3$) | 1 | 196 | 121.2 (364.0) | 47.3 | 264.2 | 3184.5 | <0.001 | <0.001 |
| | 0 | 392 | 19.2 (42.3) | 7.3 | 59.0 | 451.2 | | |
| $TSP - PM_{10}$ ($\mu g/m^3$) | 1 | 196 | 134.2 (747.7) | 0 | 246.7 | 7392.3 | <0.001 | <0.001 |
| | 0 | 392 | 15.4 (178.5) | 0 | 43.0 | 3509.3 | | |
| BC ($\mu g/m^3$) | 1 | 196 | 0.8 (0.8) | 0.6 | 2.3 | 3.6 | 0.005 | 0.001 |
| | 0 | 392 | 0.6 (0.5) | 0.5 | 1.5 | 2.7 | | |

| ***Location 8*** | **Incident** | **Count** | **Mean (Std)** | **Median** | **95th percentile** | **Max** | **p-value (MW)** | **p-value (KS)** |
|---|---|---|---|---|---|---|---|---|
| $PM_{2.5} - PM_1$ (μg/m³) | 1 | 196 | 0.5 (0.2) | 0.4 | 0.9 | 1.8 | <0.001 | <0.001 |
| | 0 | 392 | 0.4 (0.3) | 0.4 | 1.1 | 1.6 | | |
| $PM_{10} - PM_{2.5}$ (μg/m³) | 1 | 196 | 14.9 (14.7) | 9.7 | 41.0 | 72.5 | 0.029 | 0.003 |
| | 0 | 392 | 8.8 (11.5) | 3.9 | 32.3 | 71.3 | | |
| $TSP - PM_{10}$ (μg/m³) | 1 | 196 | 7.1 (23.8) | 0 | 59.5 | 198.4 | 0.046 | 0.260 |
| | 0 | 392 | 3.9 (16.3) | 0 | 31.7 | 162.9 | | |
| BC (μg/m³) | 1 | 196 | 0.2 (0.5) | 0.1 | 0.9 | 1.3 | 1 | 1 |
| | 0 | 392 | 0.5 (0.4) | 0.5 | 1.1 | 1.8 | | |

**Table S4.** Summary of Factor 3 intensity above percentile-based thresholds. *Exceedance events* are counts of continuous episodes (including single-minute episodes) during which intensity remains above the threshold. *Exceedance rate* is the percentage of one-minute observations above the threshold. *Duration* is the length of each exceedance event in minutes.

| Location | Threshold | Exceedance events (count) | Exceedance events /Hour (count) | Exceedance rate (%) | Mean duration per event (min) | Max duration per event (min) | Mean total duration /Day (min) |
|---|---|---|---|---|---|---|---|
| 1 | 90th | 12555 | 4.765 | 12.954 | 1.631 | 159 | 186.544 |
| | 95th | 7751 | 2.942 | 6.877 | 1.403 | 55 | 99.033 |
| | 99th | 2025 | 0.769 | 1.647 | 1.286 | 23 | 23.718 |
| 2 | 90th | 6709 | 3.655 | 9.206 | 1.511 | 86 | 132.563 |
| | 95th | 3555 | 1.936 | 4.312 | 1.336 | 22 | 62.098 |
| | 99th | 508 | 0.277 | 0.621 | 1.346 | 16 | 8.942 |
| 5 | 90th | 2295 | 3.740 | 7.735 | 1.241 | 18 | 111.380 |
| | 95th | 1071 | 1.745 | 3.387 | 1.164 | 16 | 48.768 |
| | 99th | 87 | 0.142 | 0.293 | 1.241 | 6 | 4.224 |
| 8 | 90th | 8378 | 3.422 | 7.984 | 1.400 | 155 | 114.963 |
| | 95th | 4698 | 1.919 | 3.900 | 1.219 | 17 | 56.158 |
| | 99th | 978 | 0.400 | 0.765 | 1.149 | 13 | 11.020 |

**Table S5.** Summary of Factor 2 intensity above percentile-based thresholds. *Exceedance events* are counts of continuous episodes (including single-minute episodes) during which intensity remains above the threshold. *Exceedance rate* is the percentage of one-minute observations above the threshold. *Duration* is the length of each exceedance event in minutes.

| Location | Threshold | Exceedance events (count) | Exceedance events /Hour (count) | Exceedance rate (%) | Mean duration per event (min) | Max duration per event (min) | Mean total duration /Day (min) |
|---|---|---|---|---|---|---|---|
| 1 | 90th | 3393 | 1.288 | 16.303 | 7.597 | 921 | 234.763 |
| | 95th | 1882 | 0.714 | 8.971 | 7.536 | 801 | 129.181 |
| | 99th | 546 | 0.207 | 2.061 | 5.967 | 265 | 29.674 |
| 2 | 90th | 1115 | 0.607 | 6.972 | 6.887 | 330 | 100.390 |
| | 95th | 534 | 0.291 | 2.860 | 5.899 | 214 | 41.181 |
| | 99th | 121 | 0.066 | 0.416 | 3.785 | 50 | 5.988 |
| 5 | 90th | 82 | 0.134 | 0.902 | 4.049 | 131 | 12.984 |
| | 95th | 48 | 0.078 | 0.320 | 2.458 | 14 | 4.615 |
| | 99th | 9 | 0.015 | 0.027 | 1.111 | 2 | 0.391 |
| 8 | 90th | 1177 | 0.481 | 7.768 | 9.693 | 626 | 111.855 |
| | 95th | 564 | 0.230 | 3.504 | 9.126 | 398 | 50.462 |
| | 99th | 123 | 0.050 | 0.541 | 6.455 | 142 | 7.784 |

# Sensitivity analysis

## Figure S8.

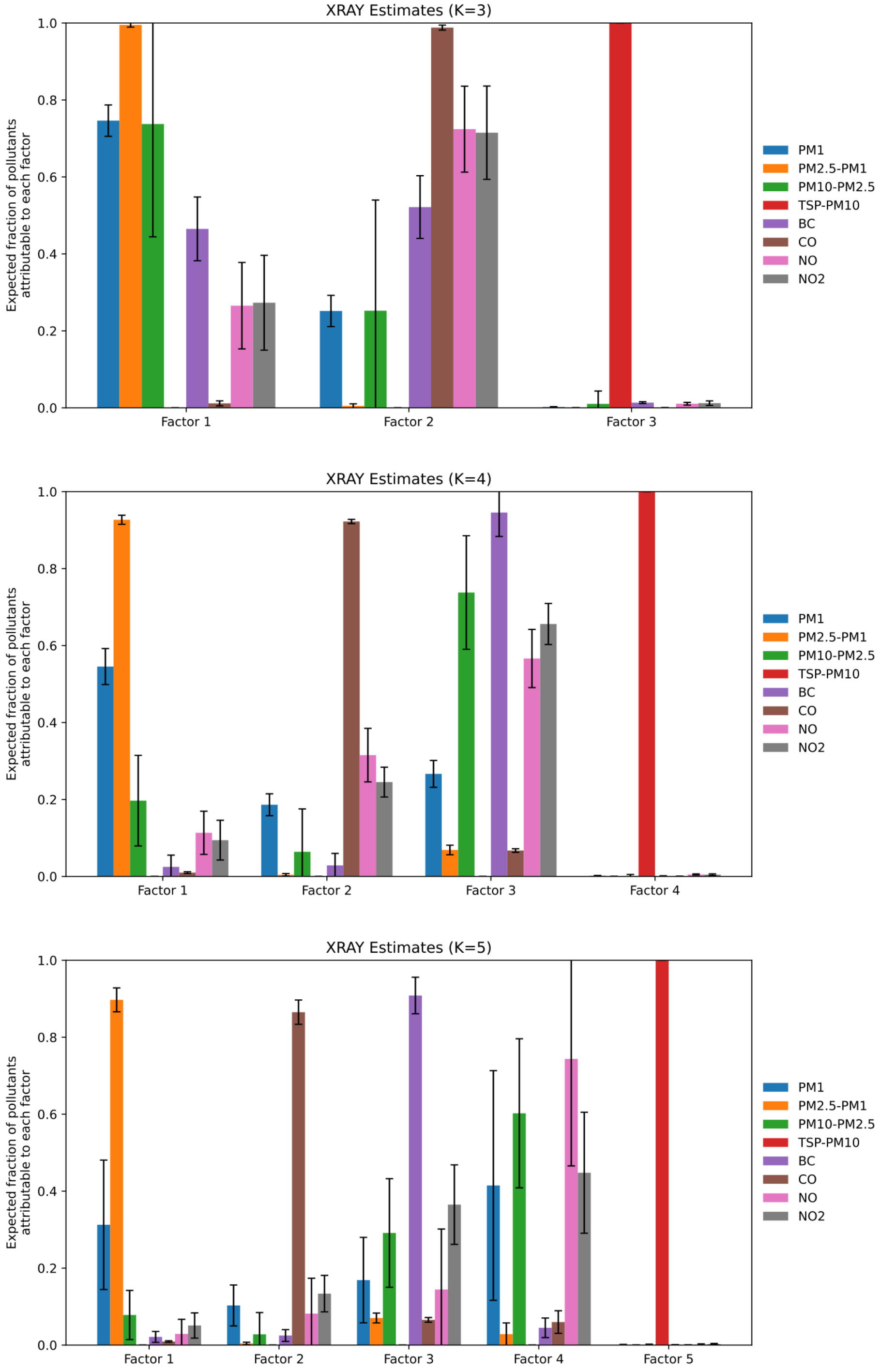

Bootstrap mean ± SE of the factor attribution matrix Φ estimated by XRAY, computed from 100 resamples for $K \in \{3,4,5\}$. For a fixed K, XRAY and GeomNMF yield highly similar attributions.

**Figure S9.**

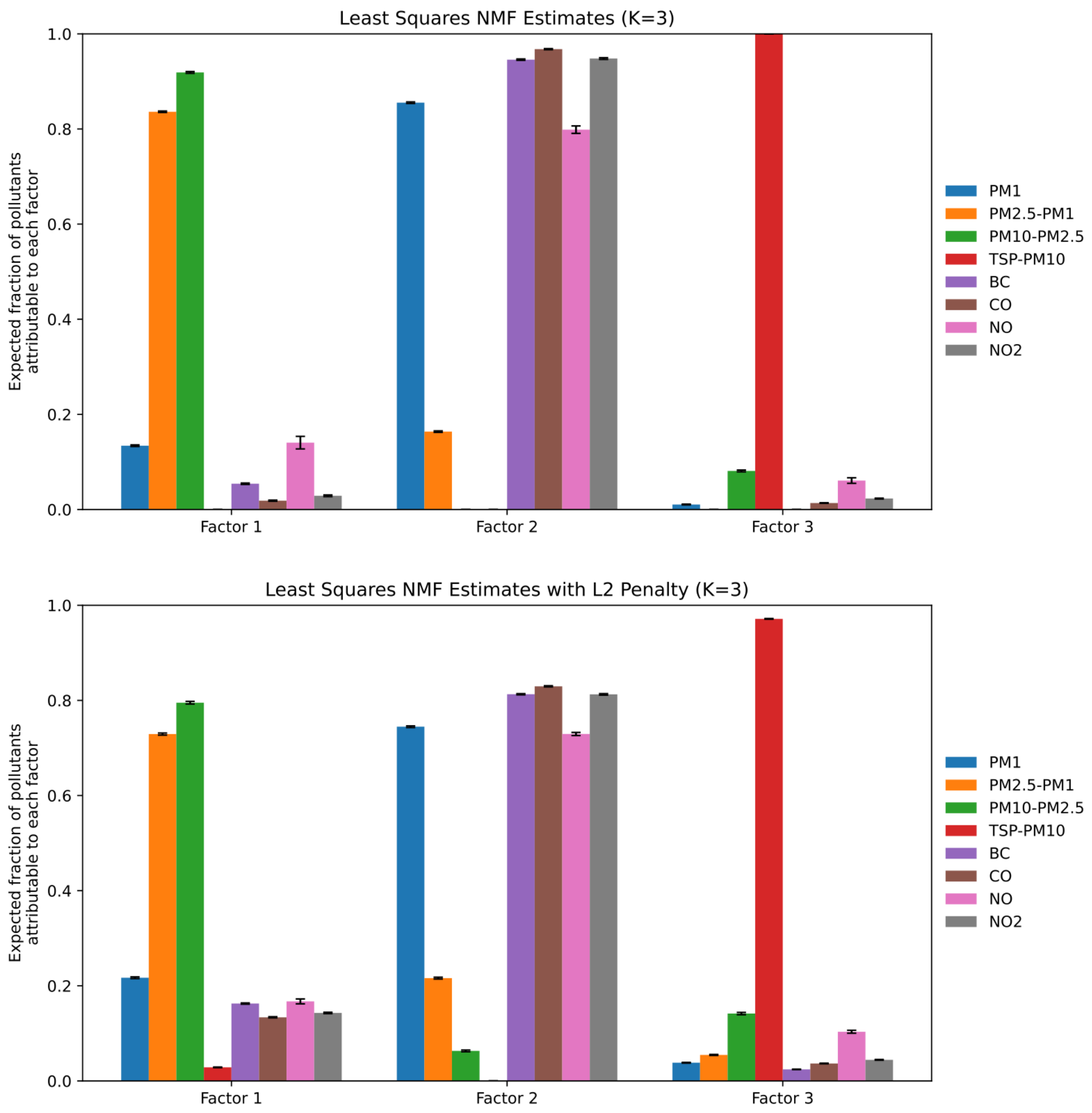

Bootstrap mean ± SE of the factor attribution matrix Φ estimated by least-squares NMF (obtained using the Python scikit-learn package) with three factors across 100 resamples. Top panel is obtained using the default setting (no regularization), while the bottom panel is obtained using L2 penalization (alpha_W = alpha_H = 0.001).

**Table S6.** Monte Carlo mean (SE) of the estimated Φ by GeomNMF ($K = 3$) from 100 randomly perturbed datasets.

| | $PM_1$ | $PM_{2.5}$-$PM_1$ | $PM_{10}$-$PM_{2.5}$ | TSP-$PM_{10}$ | BC | CO | NO | $NO_2$ |
|---|---|---|---|---|---|---|---|---|
| Factor 1 | 0.757 (0.133) | 0.987 (0.013) | 0.661 (0.315) | 0 (<0.001) | 0.315 (0.184) | 0.020 (0.018) | 0.218 (0.137) | 0.261 (0.142) |
| Factor 2 | 0.230 (0.130) | 0.011 (0.012) | 0.300 (0.298) | 0 (<0.001) | 0.663 (0.183) | 0.978 (0.018) | 0.763 (0.140) | 0.678 (0.149) |
| Factor 3 | 0.013 (0.013) | 0.001 (0.002) | 0.039 (0.091) | 1 (<0.001) | 0.021 (0.013) | 0.002 (0.001) | 0.019 (0.019) | 0.061 (0.075) |

**Figure S10.**

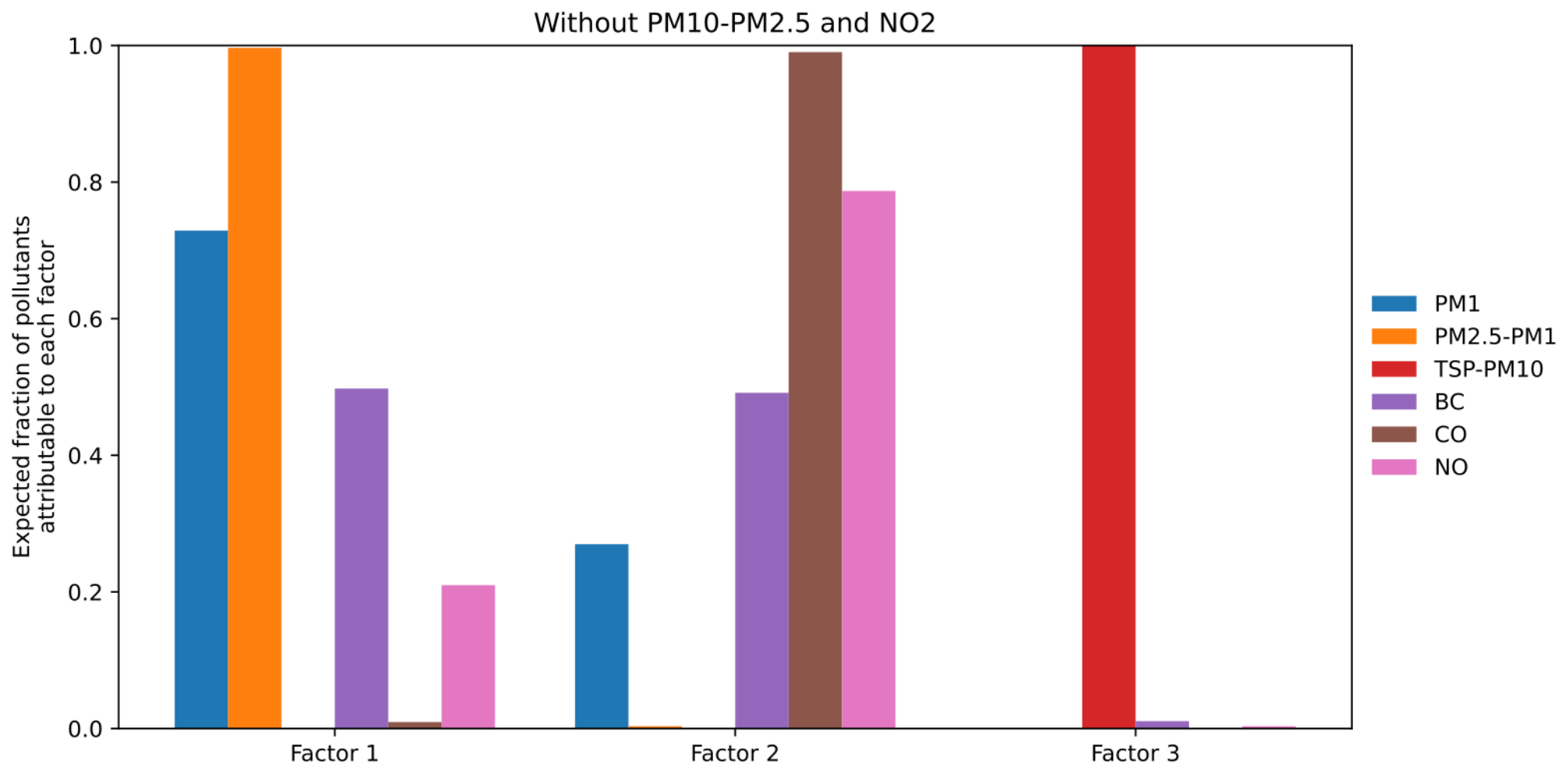

Estimated factor attribution matrix $\Phi$ by GeomNMF with $K = 3$ and $J = 6$, excluding two pollutants with the highest reported measurement uncertainty ($PM_{10}$-$PM_{2.5}$ and $NO_2$).

**Figure S11.**

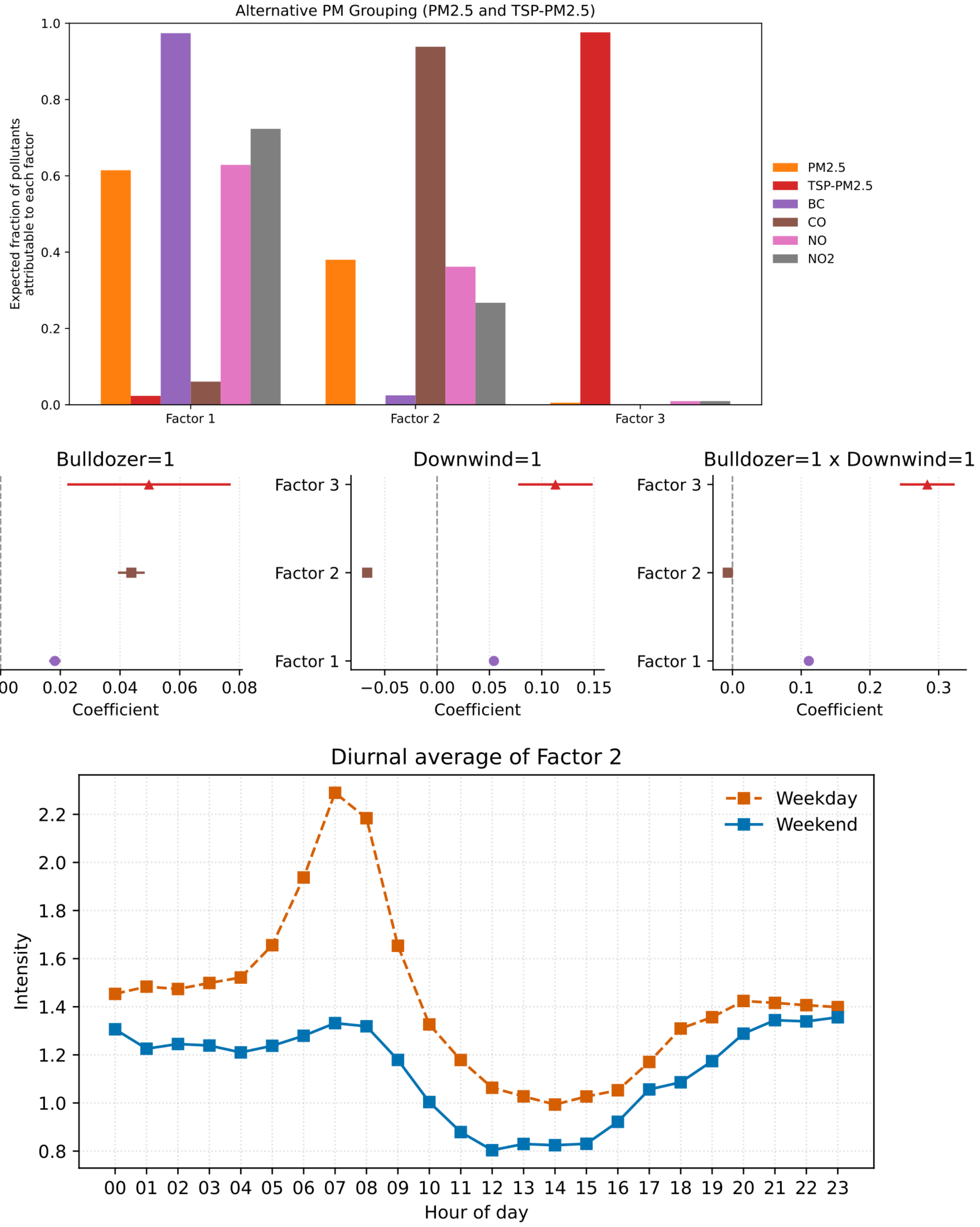

Top: Estimated factor attribution matrix Φ from GeomNMF ($K = 3$) with PM variables regrouped as $PM_{2.5}$ and TSP-$PM_{2.5}$. Middle: Regression coefficients and 95% CIs for terminal-related covariates of estimated factor intensity. Bottom: Diurnal averages of estimated Factor 2 intensity by weekday and weekend.

Figure S11 tells a consistent story with Figures 3 and 4. The CO-dominated factor (Factor 2) retains a nearly identical diurnal pattern to Factor 2 in the original grouping. Factors 1 and 3 show significantly positive coefficients for both the main effects and the main-plus-interaction effects. A minor difference is that Factor 3 in the regrouped analysis shows positive main effects, whereas the original Factor 3 was largely indifferent to those. This is expected: the regrouped TSP-$PM_{2.5}$ absorbs $PM_{10}$-$PM_{2.5}$, which was strongly positively associated with bulldozer and downwind indicators in the original grouping (see Figure S7), and this signal carries over to the new Factor 3. Results for regrouping PM as $PM_1$ and TSP-$PM_1$ are consistent with those of $PM_{2.5}$ and TSP-$PM_{2.5}$ and are therefore not shown.

**Figure S12.**

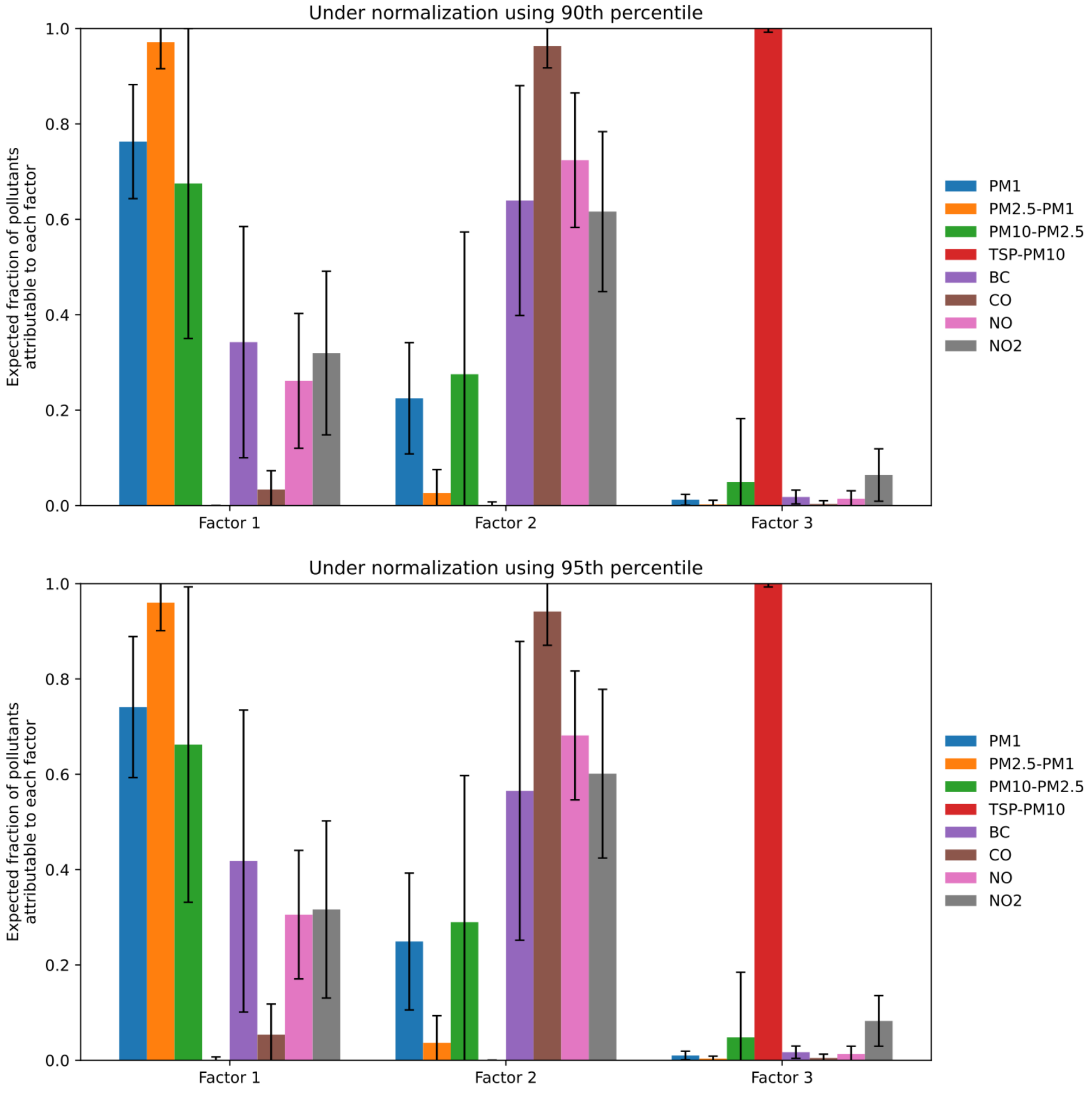

Bootstrap mean ± SE of the factor attribution matrix $\Phi$ estimated by GeomNMF ($K = 3$) after applying alternative normalization percentiles (top: 90th, bottom: 95th) across 100 resamples.

**Figure S13.**

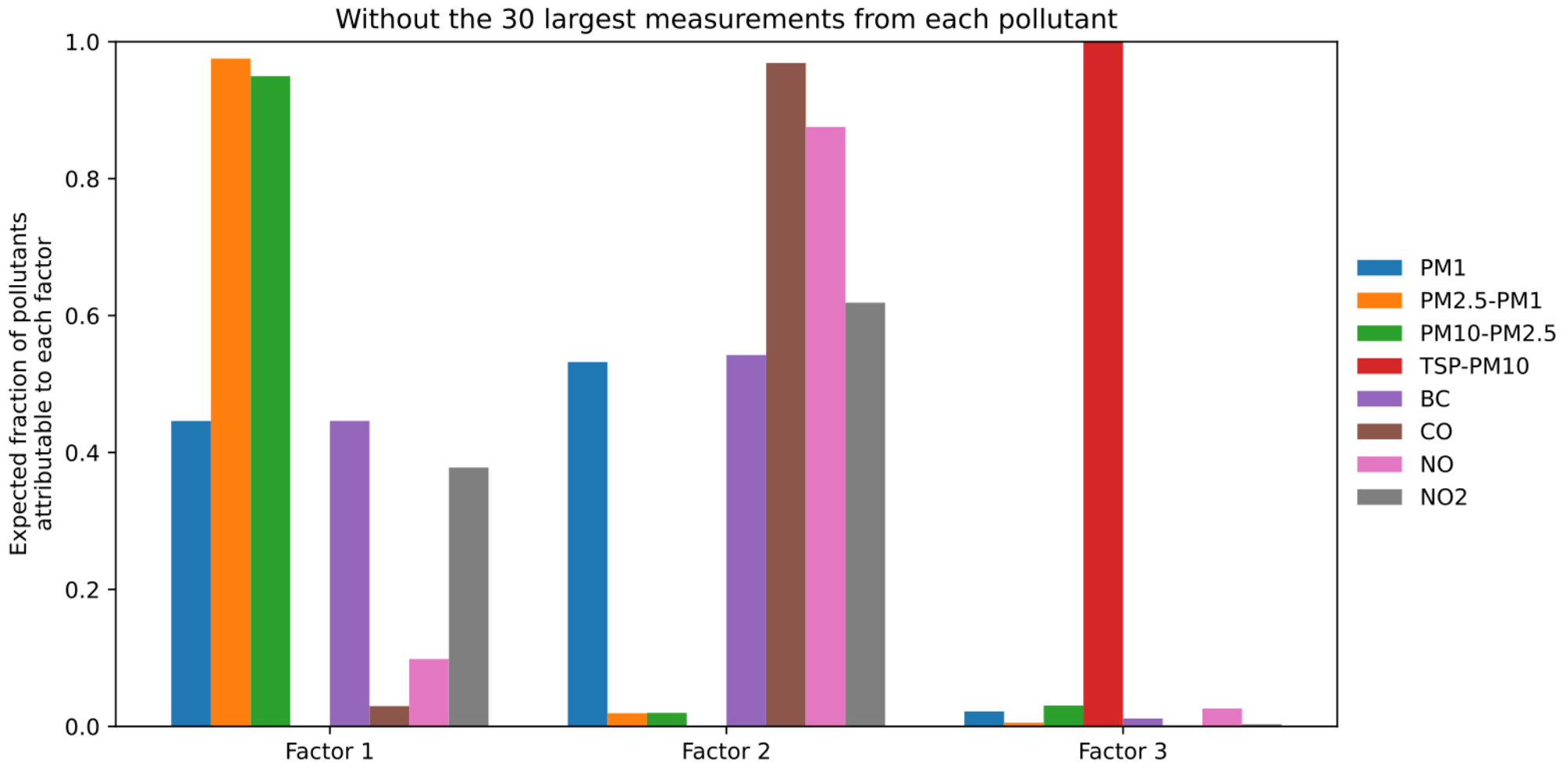

Estimated factor attribution matrix $\Phi$ from GeomNMF ($K = 3$), fit to the reduced dataset that excludes the 30 largest observations for each pollutant (225 samples, or 0.05% of the original data).